\documentclass[12pt]{iopart}

\usepackage{graphicx}
\usepackage{appendix}
\usepackage{amssymb}
\usepackage{amsmath}
\usepackage{bbold}
\usepackage{comment}
\usepackage{csquotes}
\usepackage{hyperref}
\usepackage{url}
\usepackage[usenames,dvipsnames,table]{xcolor}
\definecolor{lightgrey}{HTML}{818589}

\newcounter{mnotecount}

\expandafter\let\csname equation*\endcsname\relax
\expandafter\let\csname endequation*\endcsname\relax

\newcommand{\mnotex}[1]
{\protect{\stepcounter{mnotecount}}$^{\mbox{\footnotesize $\bullet$\themnotecount}}$ 
\marginpar{
\raggedright\tiny\em
$\!\!\!\!\!\!\,\bullet$\themnotecount: #1} }

\newcounter{mnote}

\begin{document}

\title[Hyperboloidal d'Alembert solution]{The d'Alembert solution in hyperboloidal foliations}

\author{Juan A. Valiente Kroon \& Lidia Gomes da Silva}

\address{School of Mathematical Sciences, Queen Mary, University of London, Mile End Road, London E1 4NS, UK}
\ead{j.a.valiente-kroon@qmul.ac.uk, lidiajoana@pm.me}
\vspace{10pt}

\begin{indented}
\item \today
\end{indented}

\begin{abstract}
We explicitly construct the analogue of the d'Alembert solution to the 1+1 wave equation in an hyperboloidal setting. This hyperboloidal d'Alembert solution is used, in turn, to gain intuition into the behaviour of solutions to the wave equation in a hyperboloidal foliation and to explain an apparently anomalous permanent displacement of the solution in numerical simulations discussed in the literature. 
\end{abstract}

\section{Introduction}

A hyperboloidal hypersurface in a spacetime (hyperboloid for short) is 
a spacelike hypersurface which intersects null infinity at cuts having the topology of the 2-sphere ---see e.g. \cite{CFEBook,Zen24}. Prime examples of hyperboloids are constant mean curvature (CMC) hypersurfaces in an asymptotically flat spacetime ---although, of course, these are not the only possibilities. Hyperboloidal foliations were considered in the seminal work by H. Friedrich on the semiglobal stability of the Minkowski spacetime \cite{Fri83,Fri86b}. More recently, hyperboloidal foliations have been considered as a natural way of providing gauges for the numerical evolution of the equations of linearised gravity in the context of the study of perturbations of black hole spacetimes ---see e.g. \cite{Zen11b,ZenKid10,GauVanHilBos21}. As discussed in \cite{Zen11a}, the use of hyperboloidal foliations in the study of wave equations avoids the use of unnatural boundary conditions by providing a built-in \emph{outflow boundary condition}. Alternative approaches to boundary conditions in the context of numerical simulations in General Relativity can be found in e.g. \cite{HagLau07,Lau04,Lau05,FieLau15,GroKel95}.

\medskip
An important step when working with a new type of gauge conditions, is to develop an intuition on the behaviour of some basic solutions. In the case of the wave equation in Cartesian coordinates, the basic intuition is provided by the d'Alembert solution (in the 1+1 case), the Poisson-Hadamard's solution (in the 1+2-dimensional case) and the Kirkhoff solution in the 1+3-dimensional setting. These formulae allow to compute the solutions to the wave equation in the whole of space in terms of the prescribed initial data ---see e.g. \cite{Eva98,Joh91}. 

\medskip
In the present article we restrict ourselves, for conciseness, to the discussion of the 1+1 dimensional wave equation. The more physically relevant Kirkhoff solution for the 1+3-dimensional wave equation can be recovered from the d'Alembert solution via the method of spherical decomposition means ---see e.g. \cite{Eva98}. Moreover, the 1+1-dimensional wave equations (with a potential) naturally arise from a decomposition in spherical harmonics of the solutions to the 1+3-dimensional wave equation.

Key insights provided by d'Alembert's solution into the behaviour of solutions to the 1+1-dimensional wave equation include the fact that the support of the solution on hypersurfaces of constant time (i.e. the regions where the solution is non-zero) propagates with finite speed. Moreover, if the solution had initially compact support and the initial value of its time derivative vanishes then after the front of the wave has passed, the solution returns to the trivial value (zero) ---i.e. the solution shows no permanent displacements.

\medskip
The phenomenology described in the previous paragraph is now to be contrasted with what is observed in the numerical evaluation of solutions to the 1+1 dimensional wave equation in the hyperboloidal setting \cite{eharmsmasters2012, rodrigo_pc, phdthesis-lidia}. Initial configurations with initial data of essentially compact support and vanishing time derivative do not return to the zero value after the wave front has passed. Instead, the solution settles to a constant value as it will be demonstrated by our numerical simulations implementing numerical methods developed in previous work \cite{phdthesis-lidia, da2024discotex, o2022conservativeX, da2023hyperboloidal}. In the following we refer to this behaviour as \emph{permanent displacement}. When contrasted with the Cartesian case this behaviour may appear, in first instance, puzzling. In this article we construct the analogue, for hyperboloidal foliations, of the d'Alembert solution and use it to explain the above mentioned behaviour in the numerical solutions. Further insights into the behaviour of solutions to the wave equation in the hyperboloidal wave equation can be obtained by considering various choices of initial data. Permanent displacements can also be observed in solutions to the Cartesian wave equation by choosing appropriately the initial value of the time derivative of the scalar field ---in this particular case this piece of the data appears in a term involving an integral. The novelty in the hyperboloidal setting is that the permanent displacement is produced by integral terms involving the actual value of the scalar field. Our hyperboloidal D'Alembert solution allows, in principle, to fine-tune data so as to minimise this effect.

Solutions in higher spatial dimensions, in both Cartesian and hyperboloidal settings, can be analysed by means of the method of spherical means \cite{Eva98}. This method relies on d'Alembert's formula to construct the explicit solutions to the equations. As such, analogous permanent displacements can be observed in the auxiliary variables to which d'Alembert's method is applied. However, given that the solutions in higher spatial dimensions decay (no decay occurs in 1+1 dimensions) the displacements are harder to observe in the simulations.

\begin{figure}[t]
\centering
\includegraphics[width=12cm]{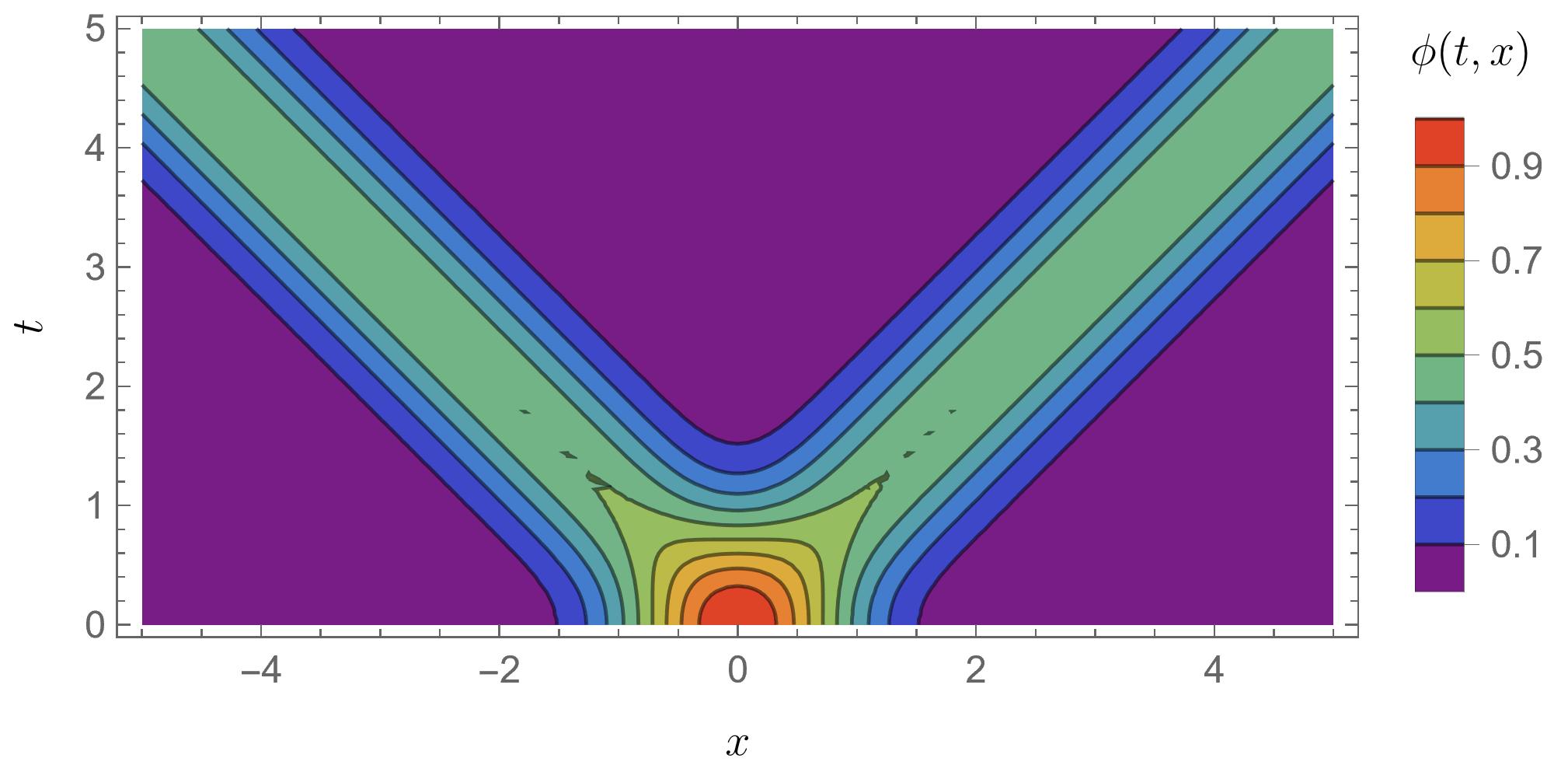}
\quad
\includegraphics[width=12cm]{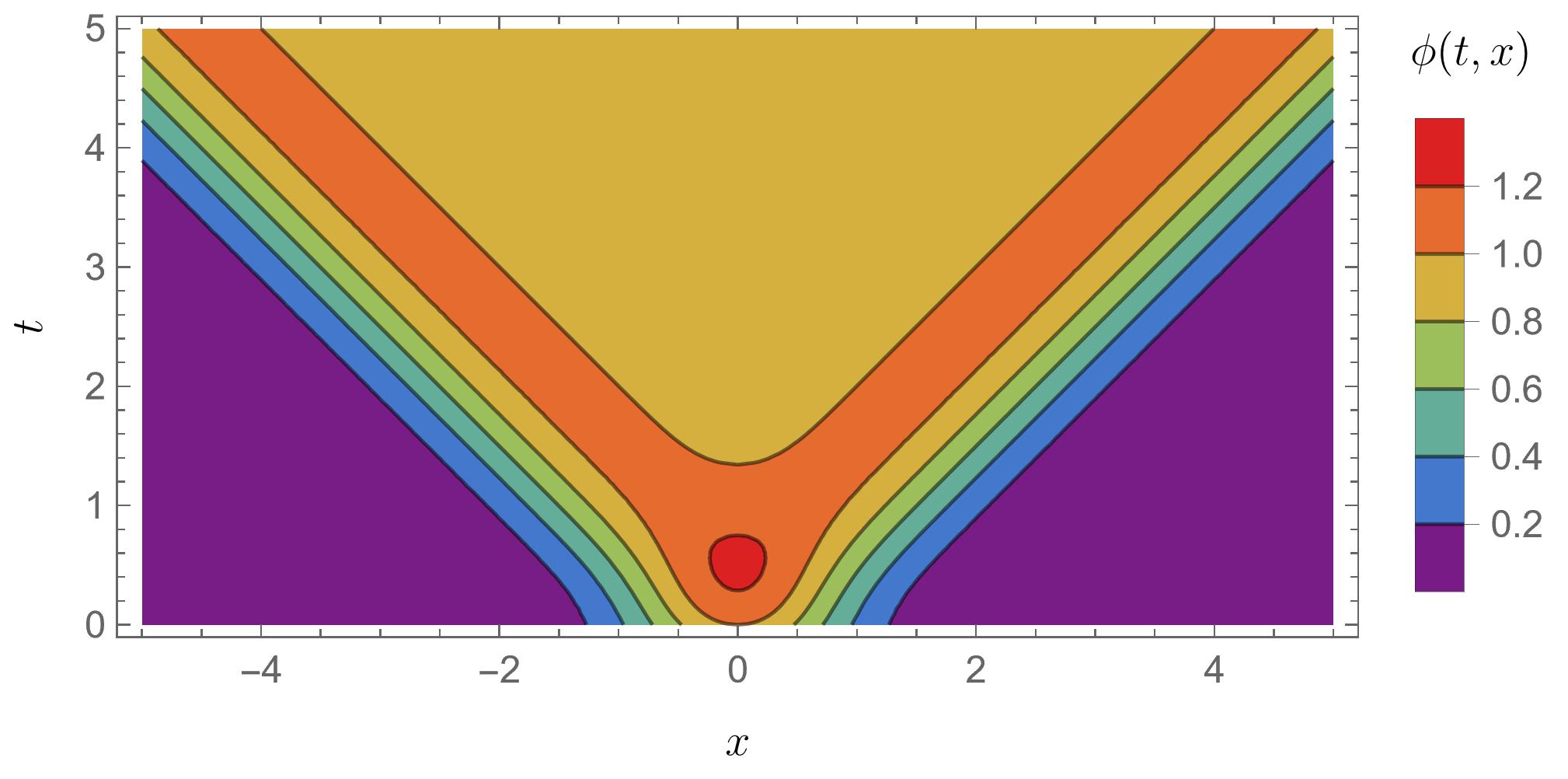}
\caption{ \textbf{Top: }contour plot of the d'Alembert solution corresponding to a choice of $f$ with the shape of a bump (i.e. of compact support) and $g=0$. This plot shows two pulses of half the size of the original bump moving in opposite directions. The support of the solution remains, at all times, compact. \textbf{Bottom:} contour plot of the d'Alembert solution for the choice $f=0$ and $g$ a bump. This choice of initial conditions generates a permanent displacement in the value of the scalar field in a compact domain.}  \label{fig:cartExactCountour}
\end{figure}

\section{The $1+1$-dimensional wave equation in Cartesian coordinates}
In this section we make some remarks regarding the properties of solutions to the wave equation in the $1+1$ and the spherically symmetric $1+3$ settings. 

\subsection{The d'Alembert solution}
We start by looking at the initial value problem for the $1+1$-dimensional wave equation on the real line. That is, we want to study the problem
\begin{subequations}
\begin{align}
     \label{IVP1}
   &\partial_t^{2} \phi - c^2 \; \partial^{2}_x \phi = 0, \qquad t\geq0, \quad x\in \mathbb{R},\\
   \label{IVP2}
    &\phi(0,x) = f(x), \\
     &\partial_t\phi(0,x)=g(x), \label{IVP3} 
\end{align}
\end{subequations}
where $f\in C^2(\mathbb{R})$ and $g\in C^1(\mathbb{R})$ are two specifiable functions over $\mathbb{R}$. Furthermore, we note here and in the rest of the manuscript we work with geometric units and thus $c =1$. 

\medskip
It is well-known that the general solution to the $1+1$-dimensional wave equation is of the form
\begin{equation}
\phi(t,x)= F(x-t) + G(x+t),
\label{GeneralSolution}
\end{equation}
where $F$, $G$ are two twice differentiable functions of a single argument. For the initial value problem \eqref{IVP1}-\eqref{IVP3} one can express the functions $F$ and $G$ in terms of the initial value problem data $f$, $g$. This gives rise to \emph{d'Alembert's solution}:
\begin{equation}
    \phi(t,x) = \frac{1}{2} \bigg( f(x-t) + f(x+t) \bigg) + \frac{1}{2} \int^{x+t}_{x-t} g(s) \mathrm{d} s.
    \label{dAlembertsFormula}
\end{equation}

\medskip
Intuition regarding formula \eqref{dAlembertsFormula} can be obtained from looking at a number of particular cases. Setting $g=0$ and choosing $f$ to have, say the shape of a \emph{bump},  gives the well-know picture of the initial bump splitting into two smaller bumps of half the size each one spreading in the opposite direction with velocity $1$. As we are in a $1+1$ setting the solution does not decay in time. If the function $f$ is of compact support then for $t>0$ the solution has also a compact support. However, the support will now consist of two disconnected parts ---see Figure \ref{fig:cartExactCountour}, top.

\medskip
An alternative situation, rarely discussed, is to set $f=0$ and set $g$ to be, again, a bump function. A particularly convenient choice of a bump function is 
\begin{equation}
g(x) = \frac{1}{1+x^2}
\nonumber
\end{equation}
which can be easily integrated. In this case d'Alembert's formula gives the solution
\begin{equation}
\phi(t,x) = \frac{1}{2}\bigg( \arctan (x+t) - \arctan(x-t) \bigg).
\nonumber
\end{equation}
Observe that for a fixed $x$, one has that $\phi(t,x)\rightarrow 1$ as $t\rightarrow \infty$. Consequently, the support of $\phi$ grows as $t$ increases. This situation is illustrated in Figure \ref{fig:cartExactCountour}, bottom.

\medskip
\noindent
\textbf{Remark.} In view of the linearity of the wave equation a generic solution is a combination of the two effects discussed previously.

\section{The wave equation in the hyperboloidal setting}

\begin{figure}[t]
    \centering
    \includegraphics[width=80mm]
    {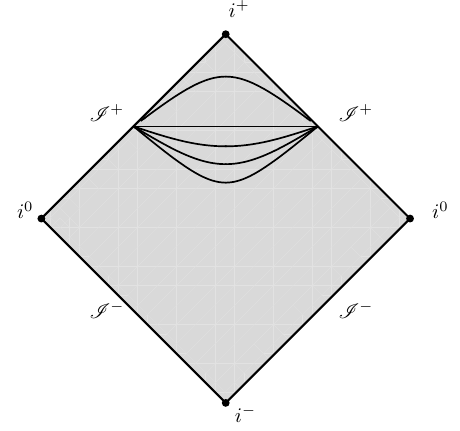}
    \caption{Examples of hyperboloids in the Minkowski spacetime corresponding, from top to bottom, to the choices of $S=7$, $S=1$, $S=\frac{1}{2}$, $S=\frac{1}{4}$  and $S=\frac{1}{10}$ in the height function $h$ as given by equation \eqref{HeightFunction}. The hyperboloids are hypersurfaces of constant mean curvature given by $K=3/\sqrt{S}$ ---see \cite{CFEBook}. The Penrose diagram is quantitatively correct in the sense that it is the plot of the height function composed with the transformation implementing the conformal embedding of the 1+1 dimensional Minkowski spacetime.}
    \label{Figure:ExamplesHyperboloids}
\end{figure}

\begin{figure}[t]
    \centering
    \includegraphics[width=80mm]{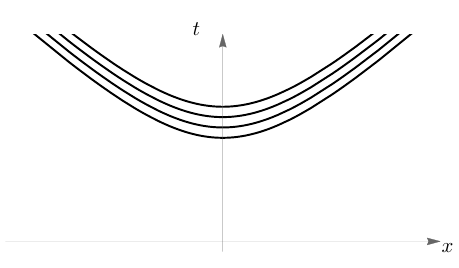}
    \quad
    \includegraphics[width=70mm]{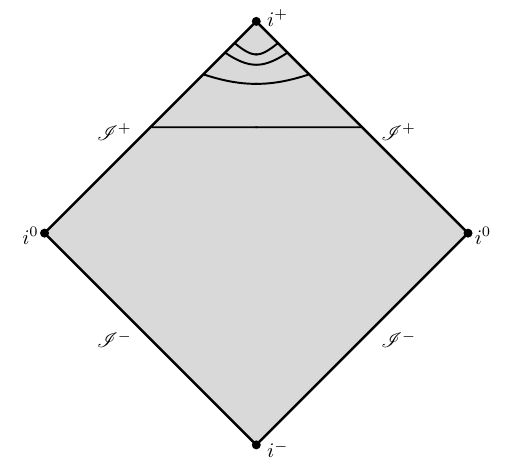}
    \caption{The hyperbolidal foliation corresponding to the choice $S=1$ for the height function $h$ given by equation \eqref{HeightFunction}. \textbf{Left:} diagram in Cartesian coordinates. \textbf{Right:} Penrose diagram. The hyperboloids correspond, from bottom to top, to the times $\tau= \{0,\,1,\,2,\,3\}$. The diagram is quantitatively correct in the sense that it is the plot of the height function composed with the transformation implementing the conformal embedding of the 1+1 dimensional Minkowski spacetime.}
    \label{fig:penrose_diagrams}
\end{figure}

Following \cite{Zen11a} we introduce coordinates $\tau$ and $\rho$ via 
\begin{eqnarray}
&& \tau \equiv t - h(x), \nonumber \\
&& x\equiv \frac{\rho}{\Omega}, \nonumber
\end{eqnarray}
where the \emph{height function} $h$ is given by
\begin{equation}
h(x) \equiv \sqrt{S^2 + x^2}
\label{HeightFunction}
\end{equation}
where $S$ is a positive number and 
\begin{equation}
 \Omega(\rho)= \frac{1}{2}\bigg(1-\frac{\rho^2}{S^2}  \bigg). 
 \nonumber
\end{equation}
It can be verified that the hypersurfaces of constant coordinate $\tau$ are hyperboloids ---i.e. spacelike hypersurfaces reaching to future null infinity; see Figure \ref{Figure:ExamplesHyperboloids}. The specific choice $S=1$ will be used in the numerical experiments discussed in this article ---see Figure \ref{fig:penrose_diagrams}.

The endpoints of the hyperboloid correspond to the sets for which $\rho=\pm S$. In particular, the $1+1$ wave equation takes the form
\begin{equation}
\partial_\tau^2 \phi + \frac{2\rho}{S}\partial_\tau\partial_\rho \phi -\Omega^2 \partial^2_\rho \phi + \frac{2S \Omega}{S^2 +\rho^2}\partial_\tau \phi + \frac{(3S^2 +\rho^2)\rho\Omega }{S^2 (S^2+\rho^2)}\partial_\rho \phi =0,
\label{HyperboloidalWaveEqn}
\end{equation}
where now $\phi=\phi(\tau,\rho)$. 
The above equation evaluated at $\rho=\pm S$ takes the form
\begin{equation}
\partial_\tau \big( \partial_\tau \mp 2 \partial_\rho \big)\phi =0.
\end{equation}
From the above factorisation it follows that one of the characteristic speeds of the system vanishes (associated to the operator $\partial_\tau$) so that the wave equation behaves, at the boundary, as an advection equation (the operator $\partial_\tau\mp\partial_\rho$) propagating information out of the system. Accordingly, the boundaries are at $\rho=\pm S$ are of an  outflow nature and, thus, no boundary conditions are required.

\medskip
We will consider the initial value problem for equation \eqref{HyperboloidalWaveEqn} on the hyperboloid with $\tau=0$. For this we prescribe, as usual
\begin{subequations}
\begin{eqnarray}
&& \phi(0,\rho) = f(\rho), \label{HypData1}\\
&& \partial_\tau \phi (0,\rho) =g(\rho), \label{HypData2}
\end{eqnarray}
\end{subequations}
where $f$ and $g$ are now thought of as functions of $\rho\in [-S,S]$ ---not to be confused with the functions with the same name used in \eqref{IVP2}-\eqref{IVP3}.

\medskip
\noindent
\textbf{Remark 1.}
In the following we will show how to adapt d'Alembert's method to obtain a formula for the solution in terms of the initial conditions.

\subsection{Construction of the solution in the hyperboidal setting}
As already discussed, the general solution to the 1+1 wave equation on the Minkowski spacetime is given by formula \eqref{GeneralSolution}. Making use of the transformation formulae between the Cartesian coordinates $(t,x)$ and the hyperboloidal coordinates $(\tau,\rho)$, one has that 
\begin{equation}
x+ t = \zeta+\tau, \qquad  x-t = \chi-\tau, 
\nonumber
\end{equation}
where we have set
\begin{equation}
\zeta \equiv \frac{S(\rho+S)}{S-\rho}, \qquad \chi \equiv \frac{S(\rho-S)}{S+\rho},
\label{TransformationCoordinates}
\end{equation}
so that the general solution to the wave equation \eqref{HyperboloidalWaveEqn} is given by
\begin{equation}
\phi(\tau,\rho)= F(\zeta +\tau) + G(\chi-\tau).
\label{GeneralSolutionHyperboloidal}
\end{equation}
Observe that $\zeta \rightarrow \infty$ as $\rho\rightarrow S$ and
$\chi \rightarrow \infty$ as $\rho\rightarrow -S$ while $\zeta(-S)=\chi(S)=0$.

\medskip
A calculation then readily shows that 
\begin{subequations}
\begin{eqnarray}
&& \phi(0,\rho) = F\left( \frac{S(\rho+S)}{S-\rho} \right)+ G\left( \frac{S(\rho-S)}{S+\rho} \right), \label{HypDataALT1}\\
&& \partial_\tau \phi(0,\rho) = F'\left(  \frac{S(\rho+S)}{S-\rho} \right) - G'\left( \frac{S(\rho-S)}{S+\rho} \right), \label{HypDataALT2}
\end{eqnarray}
\end{subequations}
where ${}'$ denotes differentiation with respect to the argument of the function. Differentiating equation \eqref{HypDataALT1} with respect to $\rho$ and taking into account the initial conditions \eqref{HypData1}-\eqref{HypData2} one obtains the system of equations
\begin{eqnarray}
&&    F'\left(  \frac{S(\rho+S)}{S-\rho} \right) - G'\left( \frac{S(\rho-S)}{S+\rho} \right) =g(\rho), \nonumber\\
&& \frac{2S^2}{(S-\rho)^2}F'\left(  \frac{S(\rho+S)}{S-\rho} \right) + \frac{2S^2}{(S+\rho)^2}G'\left( \frac{S(\rho-S)}{S+\rho} \right)=f'(\rho).\nonumber
\end{eqnarray}
The above is a linear system for
\begin{equation}
     F'\left(  \frac{S(\rho+S)}{S-\rho} \right), \qquad G'\left( \frac{S(\rho-S)}{S+\rho} \right).
     \nonumber
\end{equation}
Its solution can be readily be found to be given by 
\begin{subequations}
\begin{eqnarray}
   && F'\left(  \frac{S(\rho+S)}{S-\rho} \right) = \frac{(S^2-\rho^2)^2}{4S^2(S^2+\rho^2)}f'(\rho) + \frac{(S-\rho)^2}{2(S^2+\rho^2)}g(\rho), \label{FDashSoln}\\
   && G'\left( \frac{S(\rho-S)}{S+\rho} \right) = \frac{(S^2-\rho^2)^2}{4S^2(S^2+\rho^2)}f'(\rho) - \frac{(S-\rho)^2}{2(S^2+\rho^2)}g(\rho). \label{GDashSoln}
\end{eqnarray}
\end{subequations}
Now, re-expressing \eqref{FDashSoln} in terms of $\zeta$ as given by \eqref{TransformationCoordinates} one has that 
\begin{equation}
F'(\zeta) = \frac{S^2}{S^2+\zeta^2}g\left( \frac{S(\zeta-S)}{S+\zeta} \right)+ \frac{2S^2\zeta^2}{(S+\zeta)^2(S^2+\zeta^2)} f'\left( \frac{S(\zeta-S)}{S+\zeta}  \right).
    \nonumber
\end{equation}
Integration of this equation yields 
\begin{equation}
F(\zeta)-F(0) =\int_0^\zeta \frac{S^2}{S^2+s^2}g\left(\frac{S(s-S)}{S+s}  \right)\mathrm{d}s + \int_0^\zeta \frac{2Ss^2}{(S+s)^2(S^2+s^2)}f'\left(\frac{S(s-S)}{S+s} \right)\mathrm{d}s. 
\nonumber
\end{equation}
Using integration by parts one can remove the derivative in the second integral so that 
\begin{eqnarray}
&& F(\zeta) =F(0)+ \int_0^\zeta \frac{S^2}{S^2+s^2}g\left(\frac{S(s-S)}{S+s}  \right)\mathrm{d}s \nonumber\\
&& \hspace{2cm}+\frac{\zeta^2}{S^2+\zeta^2}f\left(\frac{S(\zeta-S)}{S+\zeta}\right)  - 2 S^2 \int_0^\zeta \frac{s}{(S^2+s^2)^2}f\left(\frac{S(s-S)}{S+s} \right)\mathrm{d}s.
\nonumber
\end{eqnarray}
Now, using \eqref{GeneralSolutionHyperboloidal} one finds, after some manipulations, that
\begin{align}
 G(\chi) &= \frac{\chi^2}{S^2+\chi^2}f\left(\frac{S(S+\chi)}{S-\chi}\right) -F(0) -\int_0^{-S^2/\chi}\frac{S^2}{S^2+s^2}g\left( \frac{S(s-S)}{S+s} \right)\mathrm{d}s \nonumber\\
&\hspace{2cm}+2S^2 \int_0^{-S^2/\chi} \frac{s}{(S^2+s^2)^2}f\left( \frac{S(s-S)}{S+s} \right)\mathrm{d}s.\nonumber 
\end{align}
Combining the above expressions using the replacements
\begin{equation}
\zeta \mapsto \zeta +\tau, \qquad \chi \mapsto \chi -\tau,
\nonumber
\end{equation}
one obtains the \emph{hyperboloidal d'Alembert formula}
\begin{eqnarray}
&& \phi(\tau,\rho) = \frac{(\zeta+\tau)^2}{S^2+(\zeta+\tau)^2} f\left( \frac{S(\zeta+\tau-S)}{S+\zeta+\tau } \right)+\frac{(\chi-\tau)^2}{S^2+(\chi-\tau)^2}f\left(\frac{S(S+\chi-\tau)}{S-\chi+\tau}\right)  \nonumber \\
&& \hspace{3cm} -2 S^2 \int^{\zeta+\tau}_{-S^2/(\chi-\tau)} \frac{s}{(S^2+s^2)^2}f\left(\frac{S(s-S)}{s+S} \right)\mathrm{d}s \nonumber\\
&& \hspace{3cm} + \int^{\zeta+\tau}_{-S^2/(\chi-\tau)} \frac{S^2}{S^2+s^2}g\left(\frac{S(s-S)}{s+S} \right)\mathrm{d}s.
\label{DAlembertHyperboloidal}
\end{eqnarray}
This is the main result in this article. It provides the solution to the wave equation in terms of the initial values of $\phi$ and $\partial_\tau\phi$ on the hyperboloid given by the condition $\tau=0$. 

\medskip
\noindent
\textbf{Remark 2.} It can be readily be verified using the relation 
\begin{equation}
\zeta = -\frac{S^2}{\chi} \nonumber
\end{equation}
that indeed $\phi(0,\rho)=f(\rho)$ and $\partial_\tau\phi(0,\rho)=g(\rho)$. Moreover, given that equation \eqref{DAlembertHyperboloidal} is manifestly of the form \eqref{GeneralSolutionHyperboloidal} allows to verify that it provides the unique solution to the initial value problem for equation \eref{HyperboloidalWaveEqn} with data given by \eqref{HypData1}-\eqref{HypData2}.


\medskip
\noindent
\textbf{Remark 3.} Equation \eqref{DAlembertHyperboloidal} is consistent with the fact that the wave equation \eqref{HyperboloidalWaveEqn} requires no boundary conditions at $\rho=\pm S$ ---i.e. that one has outflow boundary conditions. In other words, equation \eqref{DAlembertHyperboloidal} does not see the boundaries.


\subsection{Negative permanent displacements}
A peculiarity of expression \eqref{DAlembertHyperboloidal} is the integral involving $f$ ---in the third line of the equation. For ease of reference let
\begin{equation}
I(\tau,\rho) \equiv -2 S^2 \int^{\zeta+\tau}_{-S^2/(\chi-\tau)} \frac{s}{(S^2+s^2)^2}f\left(\frac{S(s-S)}{s+S} \right)\mathrm{d}s,
\nonumber
\end{equation}
with the understanding that $\chi$ and $\zeta$ are functions of $(\tau,\rho)$ via the transformation \eqref{TransformationCoordinates}. As $\tau\geq 0$, it follows that the lower integration limit in the above integral is always positive for fixed $\chi$ if $\tau$ is sufficiently large. 

\medskip
The effect of $I(\tau,\rho)$ can be better appreciated by choosing data such that $g(\rho)=0$. In this case, if $f$ is chosen to be in the form of a non-negative ``bump" (of say, compact support) then, for fixed $\rho$, 
provides a negative, monotonously decreasing contribution to the value of $\phi(\tau,\rho)$. In particular, one has that 
\begin{equation}
\lim_{\tau\rightarrow \infty} I(\tau,\rho)= -2 S^2 \int^{\infty}_{0} \frac{s}{(S^2+s^2)^2}f\left(\frac{S(s-S)}{s+S} \right)\mathrm{d}s<0.
\label{lim_tail}
\end{equation}
The integral is finite if $f$ is of compact support. Moreover, one has that its value is independent of the value of $\rho$. In fact, under the above assumptions on $f$ one has that
\begin{equation}
\phi(\infty) \equiv \lim_{\tau\rightarrow \infty} I(\tau,\rho) <0,
\nonumber
\end{equation}
as
\begin{eqnarray*}
&&\hspace{-2cm}\lim_{\tau\rightarrow \infty} \left(\frac{(\zeta+\tau)^2}{S^2+(\zeta+\tau)^2} f\left( \frac{S(\zeta+\tau-S)}{S+\zeta+\tau } \right)+\frac{(\chi-\tau)^2}{S^2+(\chi-\tau)^2}f\left(\frac{S(S+\chi-\tau)}{S-\chi+\tau}\right)\right) \\
&& \hspace{3cm}= f(S) + f(-S),
\end{eqnarray*}
which vanishes for a function $f$ of support strictly contained in $[-S,S]$. Expression \eqref{lim_tail} explains the results observed in the numerical simulations of \cite{eharmsmasters2012,rodrigo_pc, phdthesis-lidia}. That is, it shows that the solution exhibits a permanent displacement for late times. A particular example of this phenomenon will be discussed in the next section. 


\section{Numerical experiments}
\begin{figure}
    \centering
    \includegraphics[width=75mm]{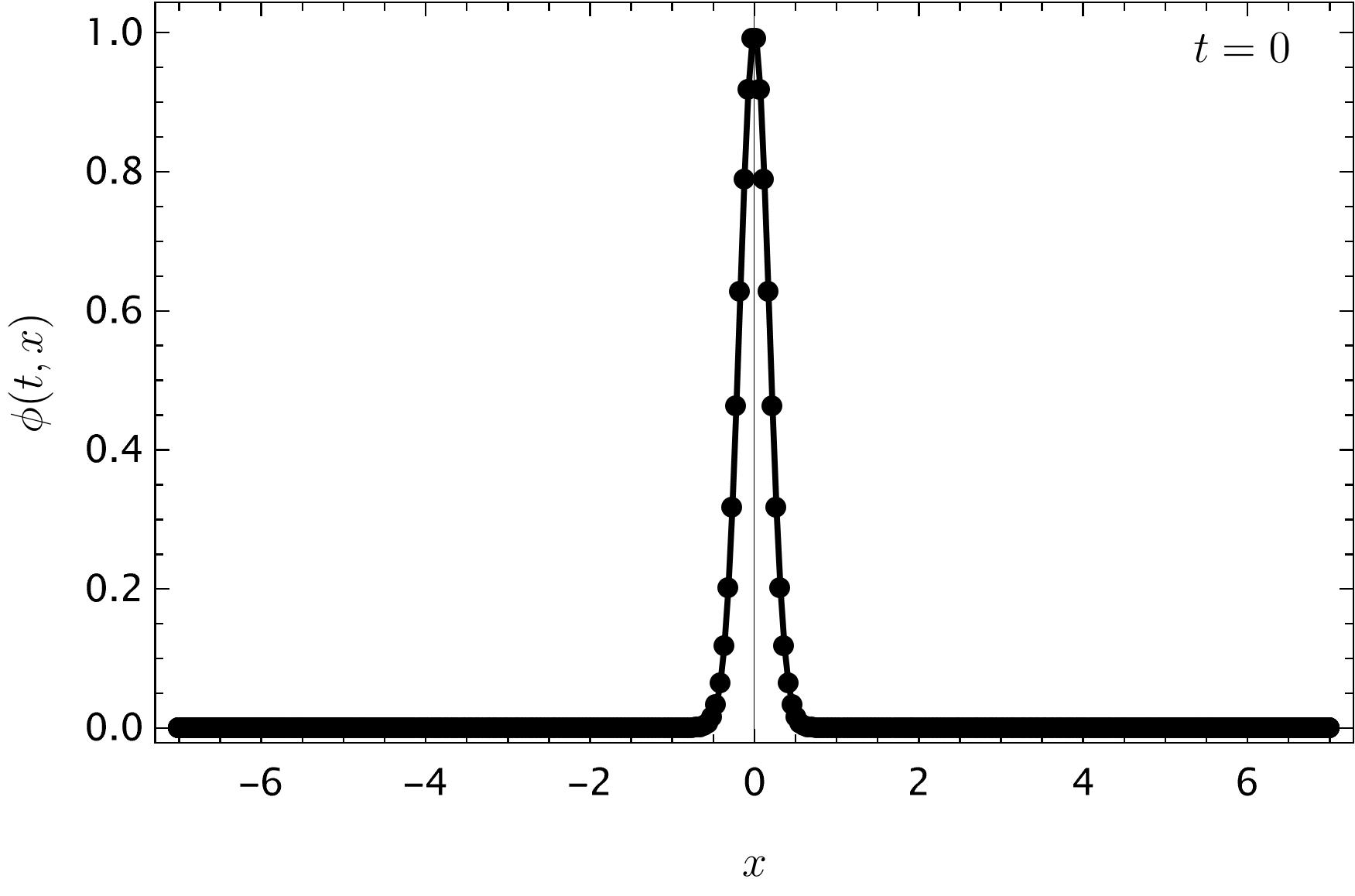}
    \quad
    \includegraphics[width=75mm]{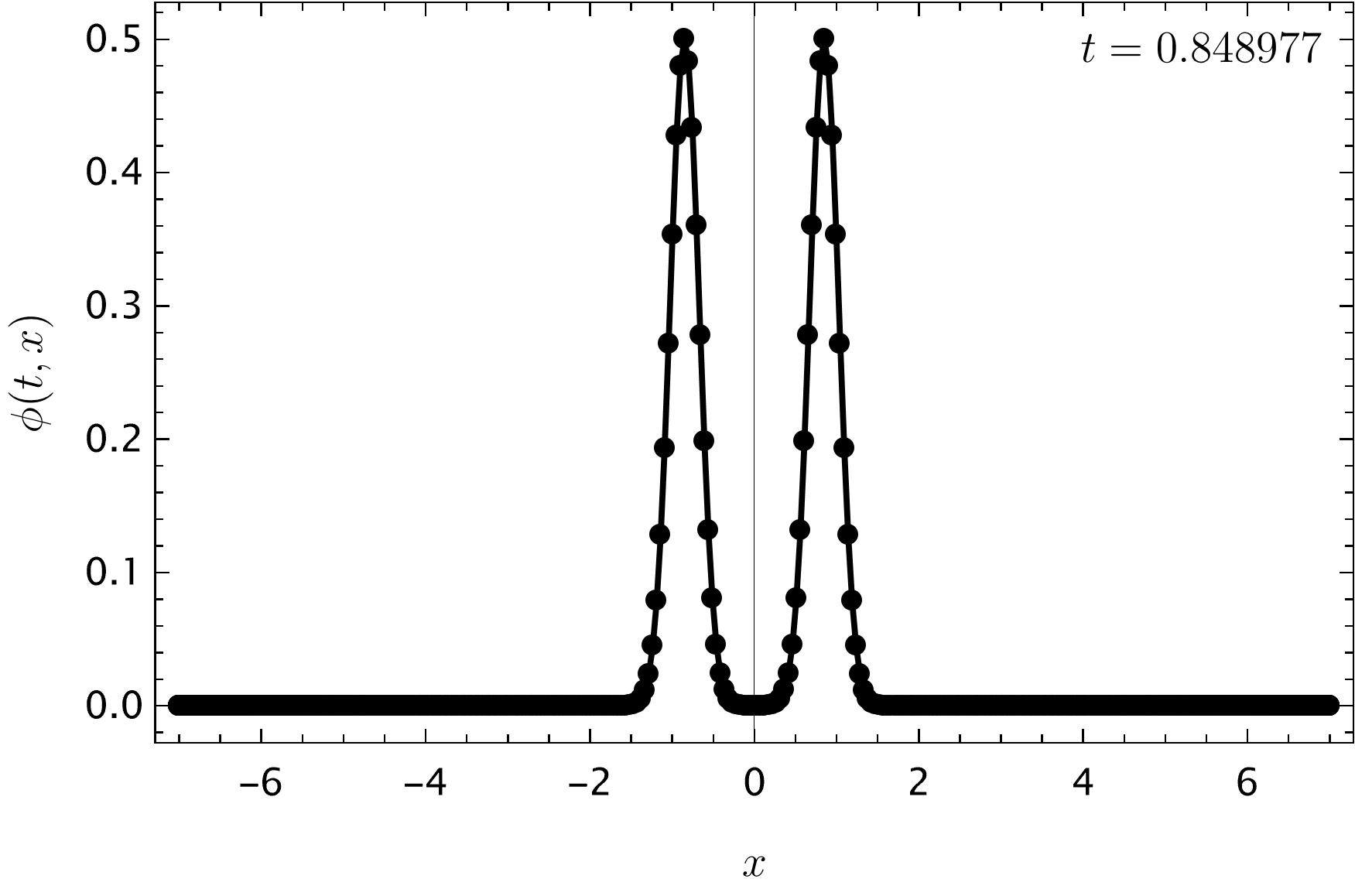}
    \quad
    \includegraphics[width=75mm]{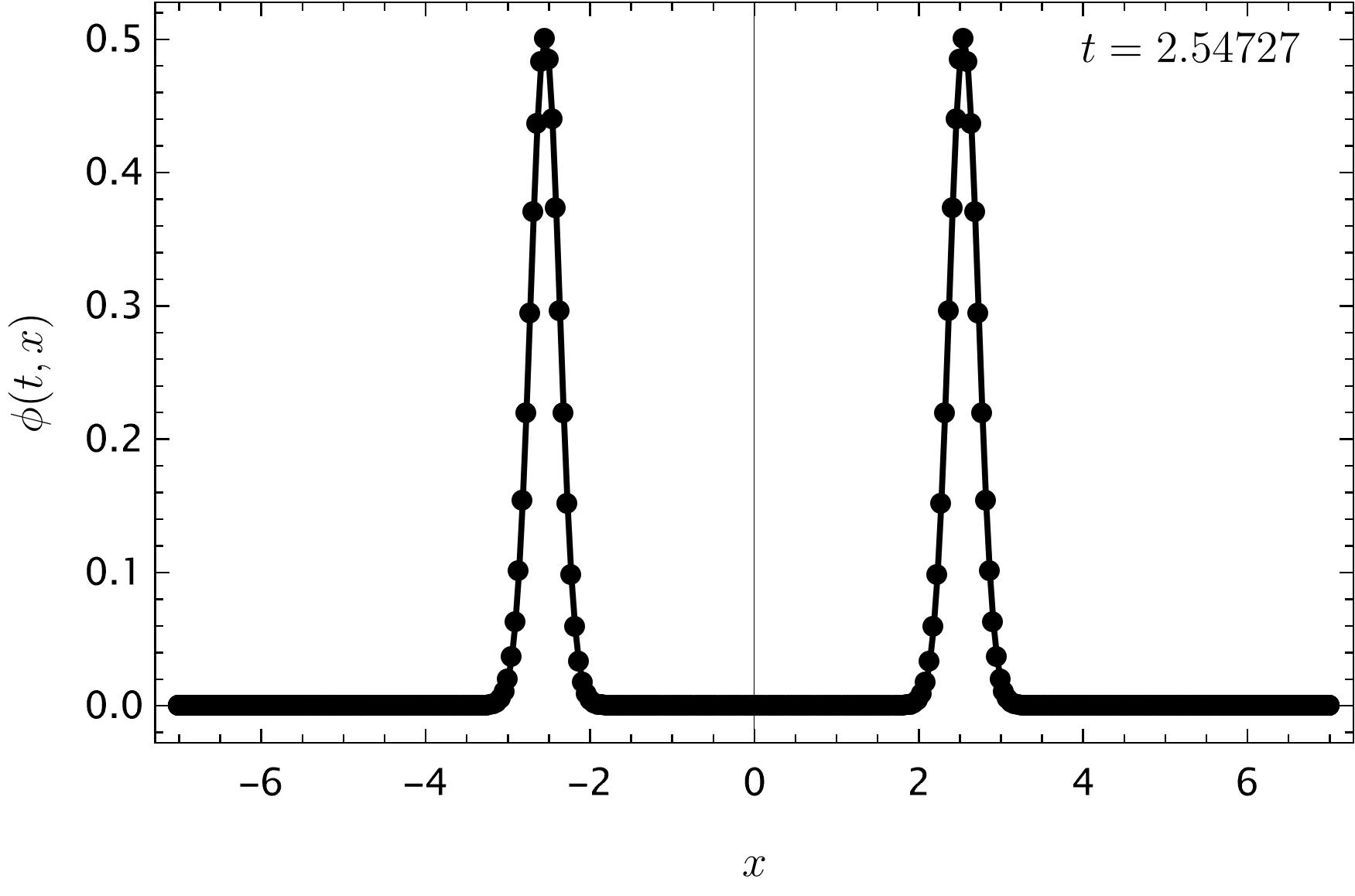}
    \quad
    \includegraphics[width=75mm]{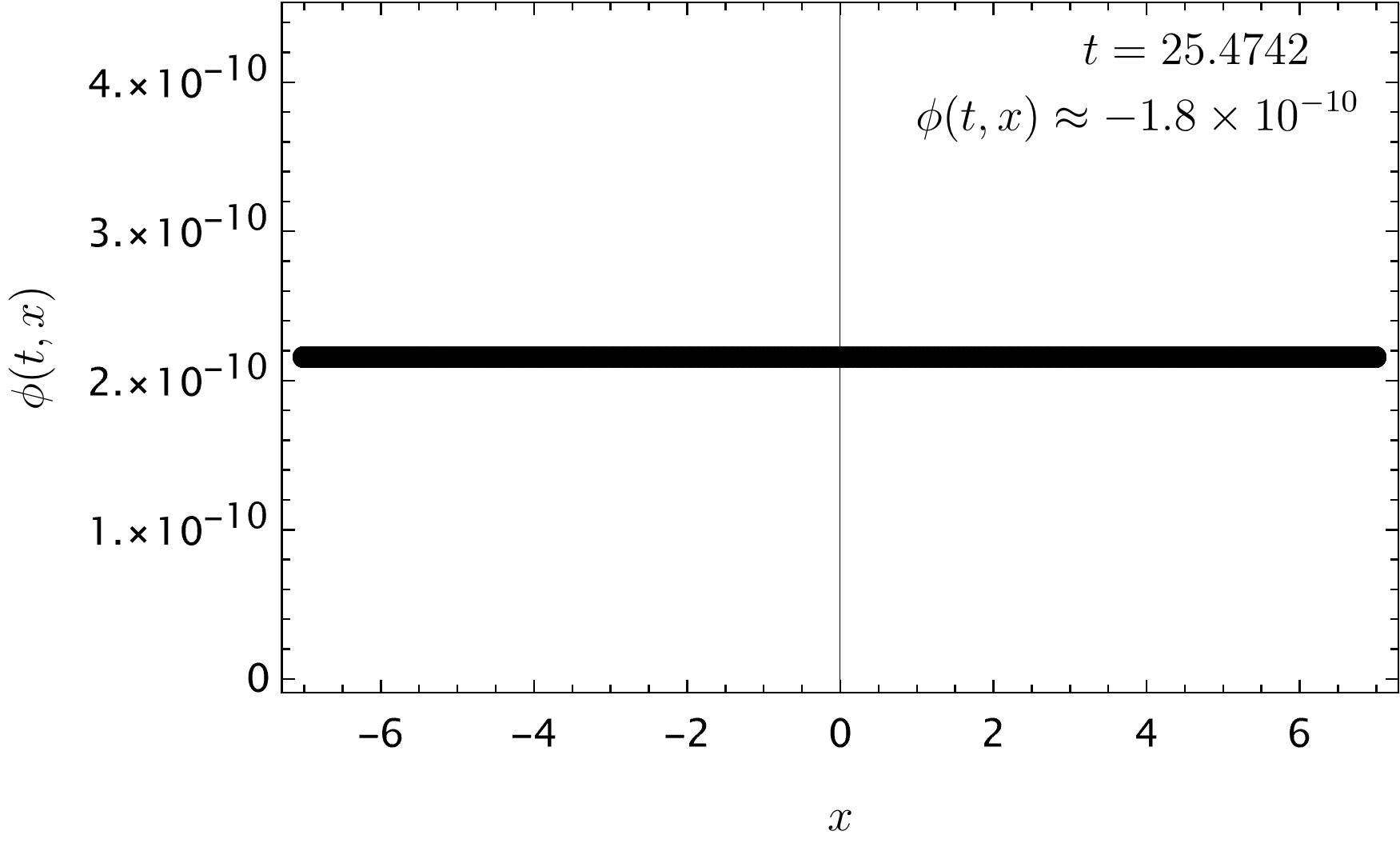}
    \caption{Numerical evolution of the \texttt{1+1D} wave equation defined by equation (\ref{IVP1}) solved numerically through the implementation of \texttt{IMTEX} Hermite time-integration scheme of order-4 with open boundary conditions as defined in equations \eqref{chap3_bcs_outgoingright}- ~\eqref{chap3_bcs_outgoingleft}. As expected outflow is observed, the initial pulse splits, travelling in opposite directions with equal field magnitudes until the solution fully leaves the numerical domain (last plot on the right).}
    \label{fig:wave_tx_evolution}
\end{figure}

\begin{figure}
    \centering
    \includegraphics[width=12cm]{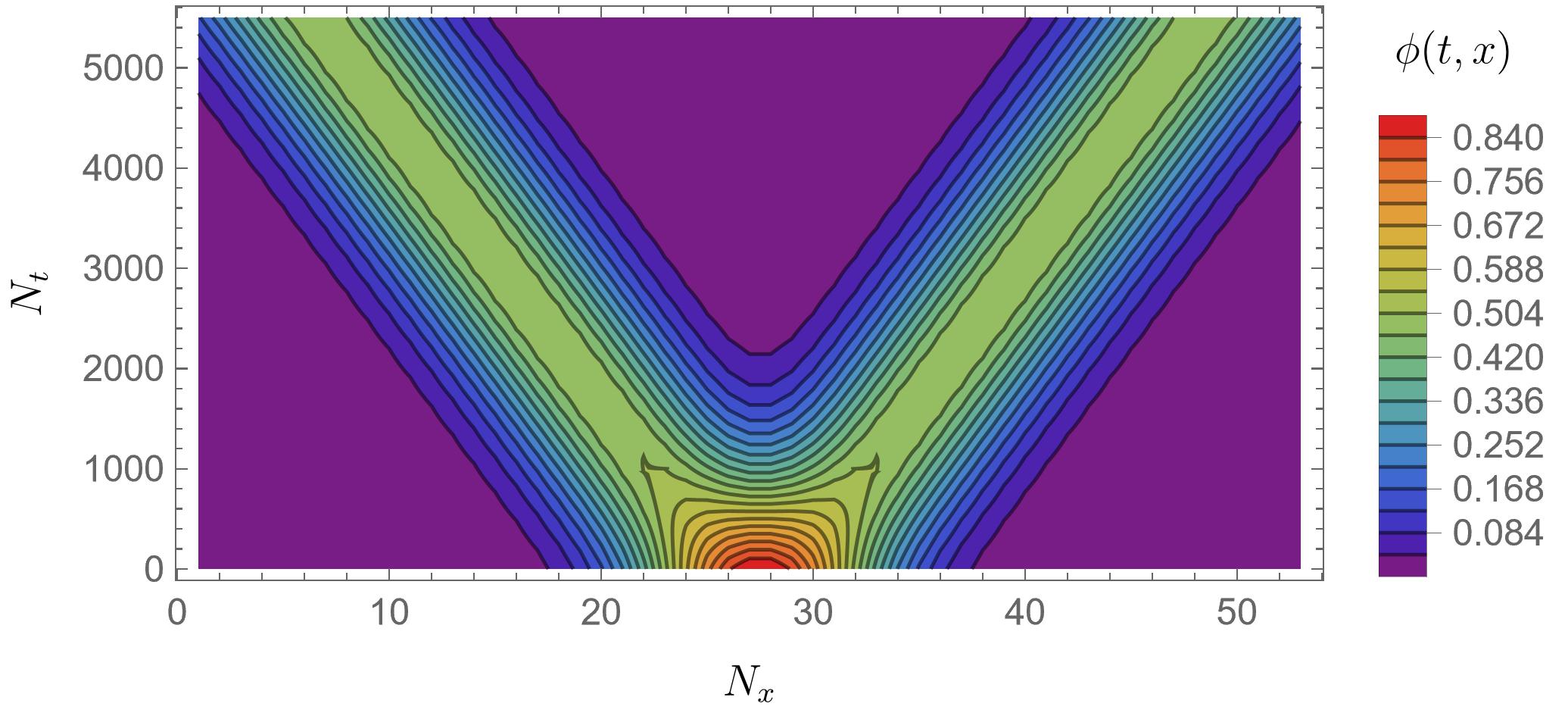}
    \caption{Contour plot further corroborating observations in Fig.(\ref{fig:wave_tx_evolution}). Initially the gaussian pulse has maximum amplitude (highlighted in red) it then splits with equal amplitudes as the simulation runs its course. Outflow is observed, and by the end of the simulation only numerical residue is detected as expected (highlighted by the colour purple).}
    \label{fig:wave_tr_contour}
\end{figure}
We will now verify our theoretical results numerically by solving equation (\ref{IVP1}) in both the original $(t,x)$ Cartesian coordinate chart and the hyperboloidal slice as defined by equation (\ref{HyperboloidalWaveEqn}). Our numerical framework is based on the method-of-lines recipe, where the partial differential equation is solved as a system of coupled ordinary differential equations (ODEs). We start by introducing the notation 
\begin{equation}
    \partial_{t} \textbf{U} = \textbf{L} \ \textbf{U},  \ \ \ \ \textbf{U} = \begin{pmatrix}\phi \\ \mathcal{\psi} 
    \end{pmatrix}, 
    \label{red_pdes}
\end{equation}
where the matrix evolution operator is defined as
\begin{align}
    &\textbf{L} = \begin{pmatrix} 0 & \mathbb{1} \\ \textbf{L}_{1} & \textbf{L}_{2}\end{pmatrix},
    \label{Genric_matrix_op}
\end{align}
and finally reduce the system to the reduced ODE 
\begin{equation}
    \frac{d \textbf{U}}{dt} = \textbf{L} \, \textbf{U}. 
\end{equation}

In this framework, the fields $\phi(t,x)$ and $\mathcal{\psi}(t,x)$ are evaluated on a discrete spatial grid $x = \{x_{i}\}^{N}_{i=0}$ such that $\phi(t,x)\rightarrow$ $ \boldsymbol{\phi}(t)$ and $\mathcal{\psi}(t,x)\rightarrow$ $ \boldsymbol{\mathcal{\psi}}(t)$. The components $\phi(t,x_{i}) \equiv \phi_{i}(t) $ and $\mathcal{\psi}(t,x_{i}) \equiv \mathcal{\psi}_{i}(t) $ of the vectors $\boldsymbol{\phi}(t)$ and $ \boldsymbol{\psi}(t)$ are the fields evaluated on the grid points. Generically as we demonstrated in \cite{phdthesis-lidia,  da2024discotex, o2022conservativeX, da2023hyperboloidal} we will be solving a system of coupled \texttt{2N+2} time-dependent ODEs given as
\begin{equation} \label{phi_psi_ODEs}
\frac{d}{{d{t}}}
    \begin{pmatrix}
        \phi_\imath \\
        \mathcal{\psi}_\imath
    \end{pmatrix}
= \sum\limits_{\jmath = 0}^N {{\begin{pmatrix}
0&{{\textbf{I}_{\imath \jmath}}}\\
{{\tilde{\chi}_\imath }D_{\imath \jmath}^{(2)} + {\tilde{\iota}_\imath}D_{\imath \jmath}^{(1)} + {\tilde{\mathcal{V}}_\imath }{\textbf{I} _{\imath \jmath}}}&{{\tilde{\varepsilon}_\imath }D_{\imath \jmath}^{(1)} + {\tilde{\varrho}_\imath }{\textbf{I} _{\imath \jmath}}}
\end{pmatrix}} 
    \begin{pmatrix}
        \phi_\jmath \\
        \mathcal{\psi}_\jmath
    \end{pmatrix}
},\quad \imath = 0,1,...,N
\end{equation}
where $\tilde{\chi}_{\imath} = \tilde{\chi}(x_{\imath}), \tilde{\iota}_{\imath} = \tilde{\iota}(x_{\imath}), \tilde{\mathcal{V}}_{\imath} = \tilde{\mathcal{V}}(x_{\imath}), \tilde{\varepsilon}_{\imath} = \tilde{\varepsilon}(x_{\imath}), \tilde{\varrho}_{\imath} = \tilde{\varrho}(x_{\imath})$ and $\textbf{I}_{\imath \jmath}$ is the \texttt{N+1} identity matrix, obtained in \texttt{Mathematica} for example as, $\textbf{I}$ \texttt{= IdentityMatrix[N+1]}. For the wave equation in the $(t,x)$ chart we have, $\tilde{\chi}_{\imath} = \tilde{\chi}(x_{\imath}) \rightarrow 1 $ and $\tilde{\iota}_{\imath} = \tilde{\mathcal{V}}_{\imath}= \tilde{\varepsilon}_{\imath} =  \tilde{\varrho}_{\imath} = 0$. For the hyperboloidal chart defining equation (\ref{HyperboloidalWaveEqn}) we have
\begin{subequations}
\begin{align}
    \label{ch3_sb2_diffOperators_varepsilon}
    & \tilde{\varepsilon}(\rho) = \frac{2\rho}{S}, \\
    \label{ch3_sb2_diffOperators_varrho}
    &\tilde{\varrho}(\rho) =  \frac{2S \Omega}{S^2 +\rho^2} , \\
    \label{ch3_sb2_diffOperators_chi}
    &\tilde{\chi}(\rho) = \Omega^2,\\
    \label{ch3_sb2_diffOperators_potential} 
    &\tilde{\mathcal{V}}(\rho) = 0, \\
    &\tilde{\iota}(\rho) = \frac{(3S^2 +\rho^2)\rho \Omega }{S^2 (S^2+\rho^2)}. 
    \label{ch3_sb2_diffOperators_iota}
\end{align}
\end{subequations}
For spatial discretisation in the spatial domain $x \in [a,b]$ where $a \leq x_{i} \leq b$ we use Chebyshev collocation nodes ranging from $0<i<N$ and follow the recipe given in Section 3.2.1 of \cite{da2024discotex}. Particularly, in this work we use  \texttt{Mathematica Version 13.2} to obtain the \texttt{n-th} order differentiation matrices with the command,
\begin{verbatim}
Dn = NDSolve`FiniteDifferenceDerivative[Derivative[i], x, 
DifferenceOrder -> "Pseudospectral"]@"DifferentiationMatrix"].
\end{verbatim}

\begin{figure}
    \centering
    \includegraphics[width=7.3cm]{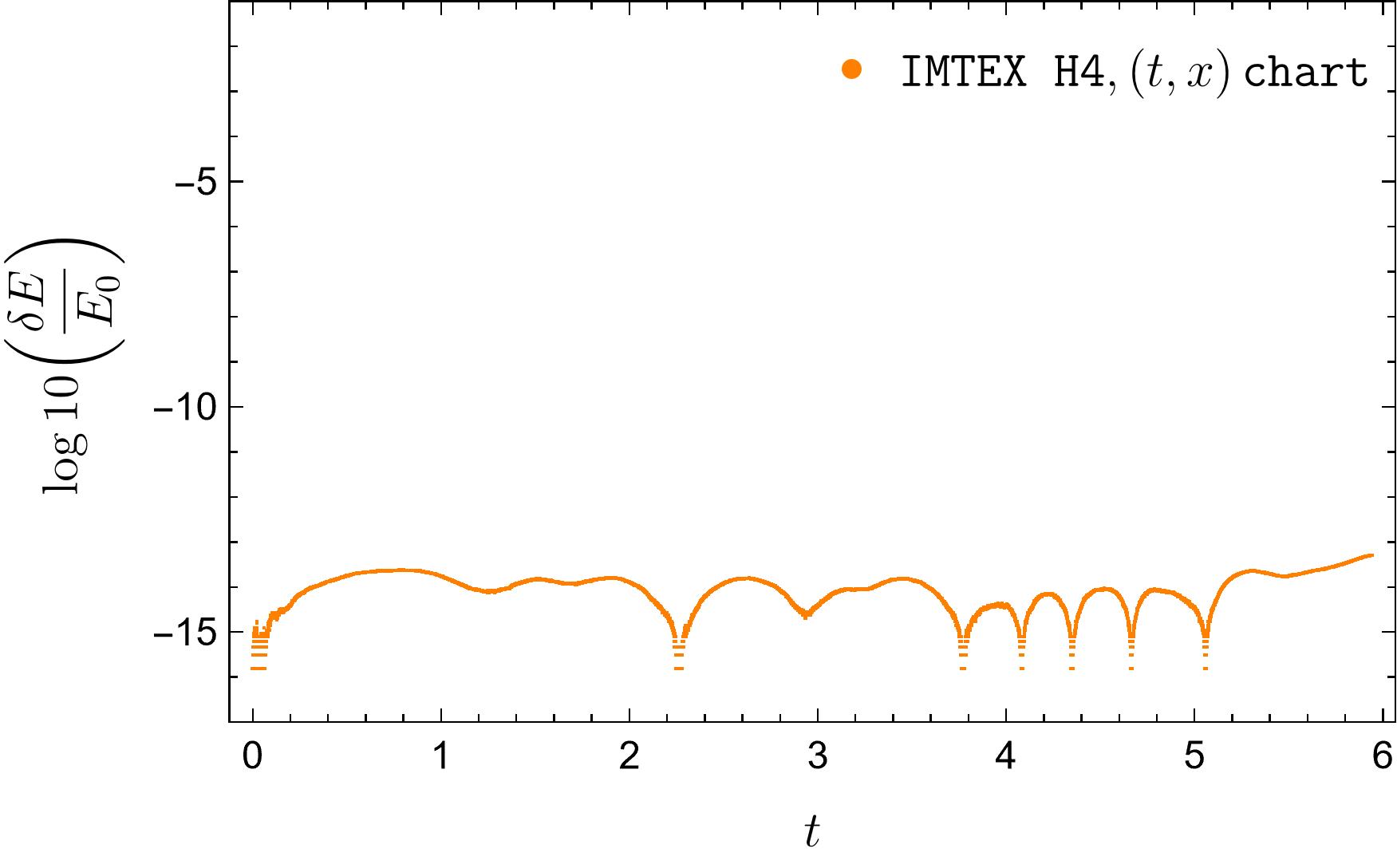}
    \quad
    \includegraphics[width=7.1cm]{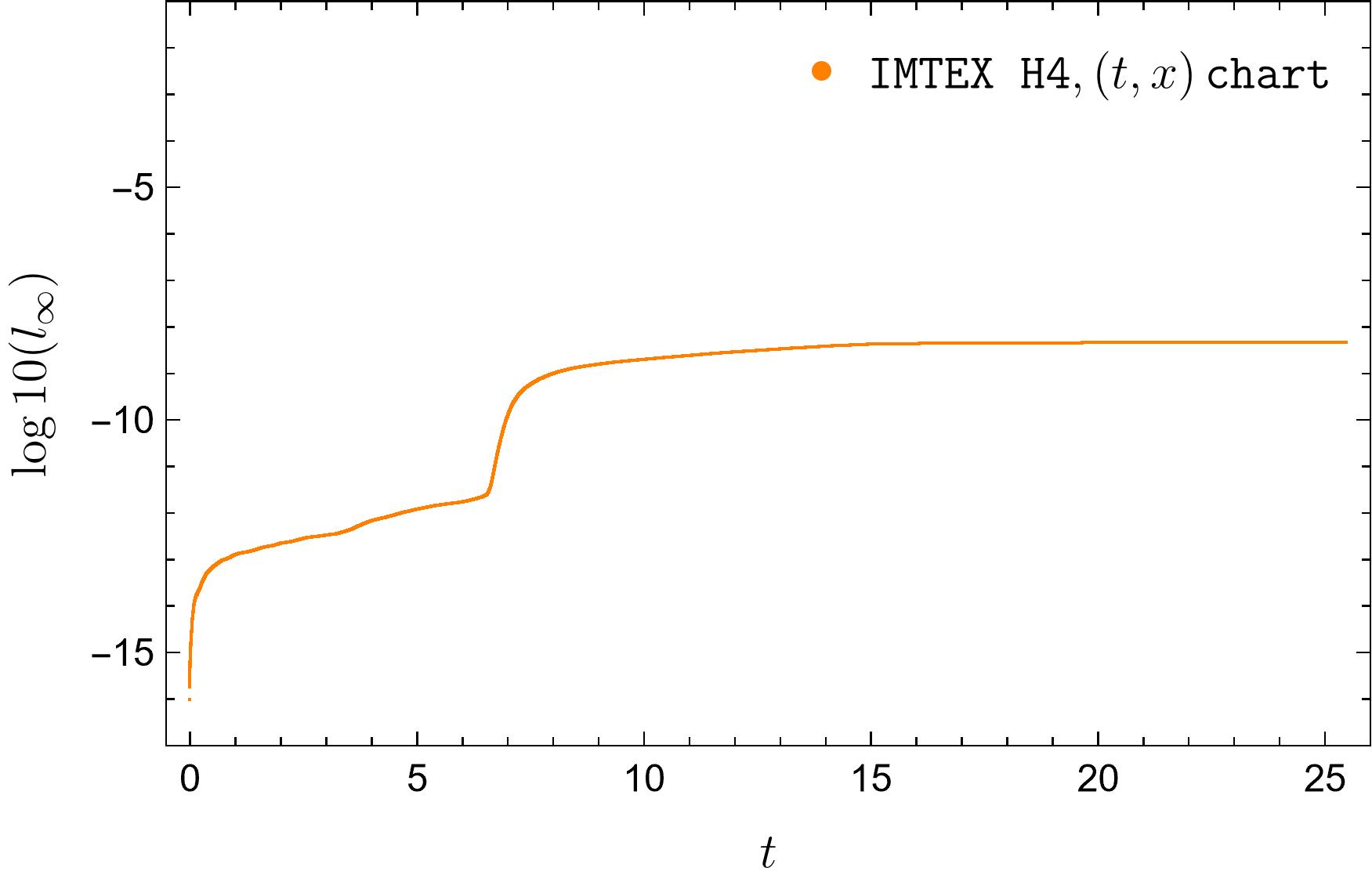}
    \caption{Numerical solution to the wave equation in the $(t,x)$ coordinate chart.  \textbf{Left}: fractional energy error of the numerical solution up to when the pulse is inside the computational domain. High accuracy is observed and energy is conserved both exactly and numerically \cite{phdthesis-lidia,da2024discotex, o2022conservativeX}. \textbf{Right:} numerical error between the d'Alembert solution given by eq.(\ref{dAlembertsFormula}) and with the numerical recipe of eq.(\ref{ch3_hornerfor_sh4} - \ref{num_TX_id_2}) in the numerical domain $x \in [-7,7]$ and $t \in [0, 25.4742]$}.
    \label{fig:wave_tr_con_energy_linf}
\end{figure}

The time-component is resolved numerically by using the implicit-turned-explicit \texttt{IMTEX} Hermite time-integration schemes \cite{phdthesis-lidia,da2024discotex, o2022conservativeX, da2023hyperboloidal}. In this work one opts for the fourth-order Hermite \texttt{IMTEX} scheme, given as 
\begin{subequations}
\begin{small}
\begin{align}
    \label{ch3_hornerfor_sh4}
    &\textbf{U}_{n+1} = \textbf{U}_{n}+ (\Delta t \; \textbf{L})  \cdot \textbf{HFH4} \cdot \textbf{U}_{n}, \\
    &\textbf{HFH4} = \bigg( \textbf{I} - \frac{\Delta t \; \textbf{L}}{2} \cdot \bigg( \textbf{I} - \frac{\Delta t \; \textbf{L}}{6}  \bigg)\bigg)^{-1}.  
    \label{ch3_hfh4}
\end{align}  
\end{small}
\end{subequations}

For details and comparisons with higher-order \texttt{IMTEX} Hermite schemes, purely implicit \texttt{IM} Hermite and purely explicit \texttt{EX} \texttt{Runge-Kutta} schemes we refer the reader to \cite{phdthesis-lidia, da2024discotex,o2022conservativeX}. 

\subsection{Numerical solution to the wave equation in the $(t,x)$ coordinate chart with outflow boundary conditions}

Having selected our numerical framework, and based on the discussion in previous chapters, we will first study the behaviour of our numerical solution in the $(t,x)$ coordinate chart with outflow boundary conditions defined as, 
\begin{subequations}
\begin{eqnarray}
    \label{chap3_bcs_outgoingright} 
   \partial_{t}\phi(t,x_{0}) =  c \  \partial_{x}\phi(t,x), \\  
   \partial_{t}\phi(t,x_{N}) = - c \ \partial_{x}\phi(t,x). 
   \label{chap3_bcs_outgoingleft}   
\end{eqnarray}
\end{subequations}
For initial data we pick, in accordance to equations (\ref{IVP2}, \ref{IVP3}),
\begin{subequations}
\begin{align}
    \label{num_TX_id_1}
    & f(x) = e^{- 16 x^{2}}, \\
    & g(x) = 0. 
    \label{num_TX_id_2}
\end{align}
\end{subequations}
\begin{figure}
    \centering
    \includegraphics[width=7cm]{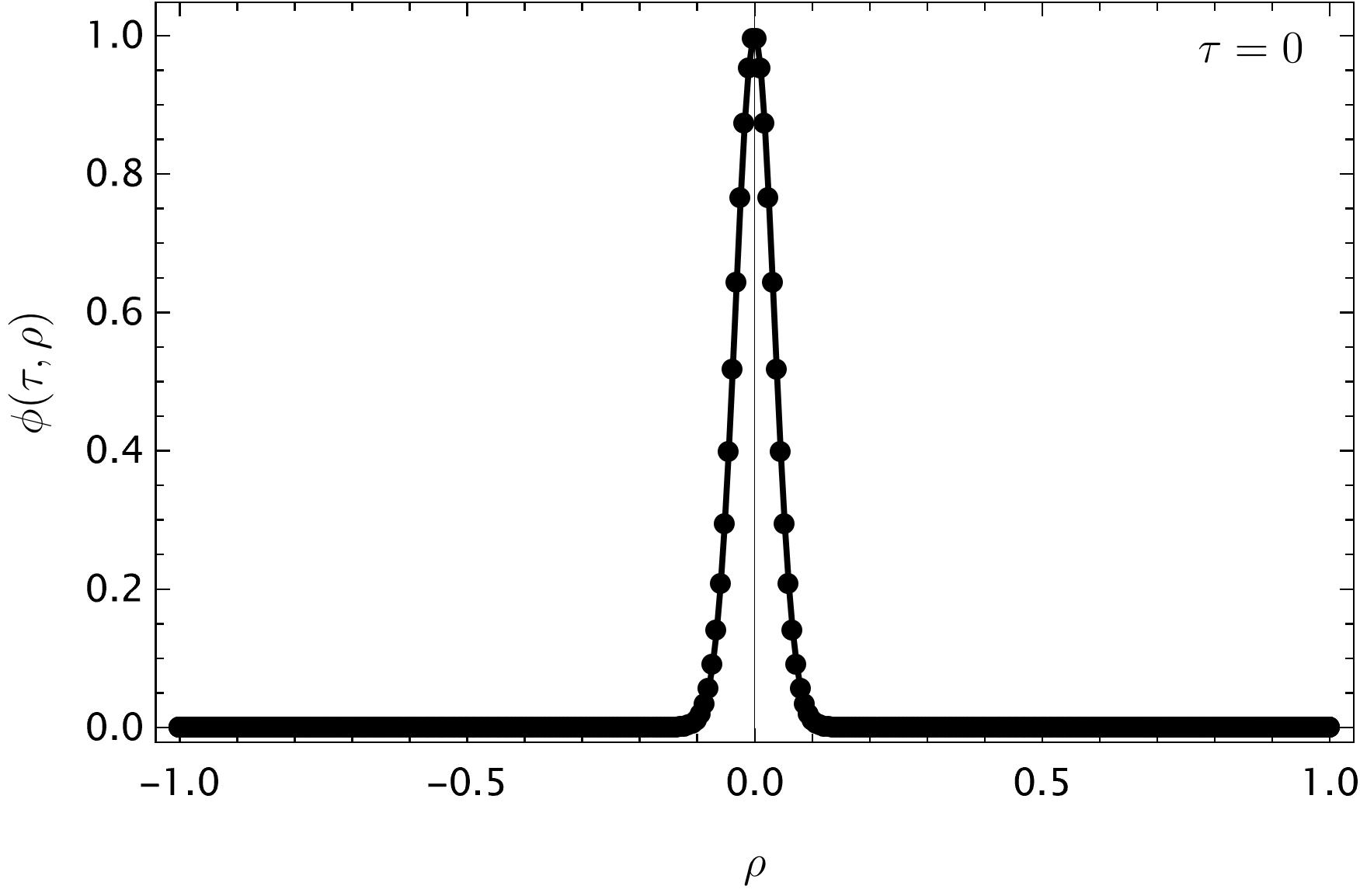}
    \quad
    \includegraphics[width=7cm]{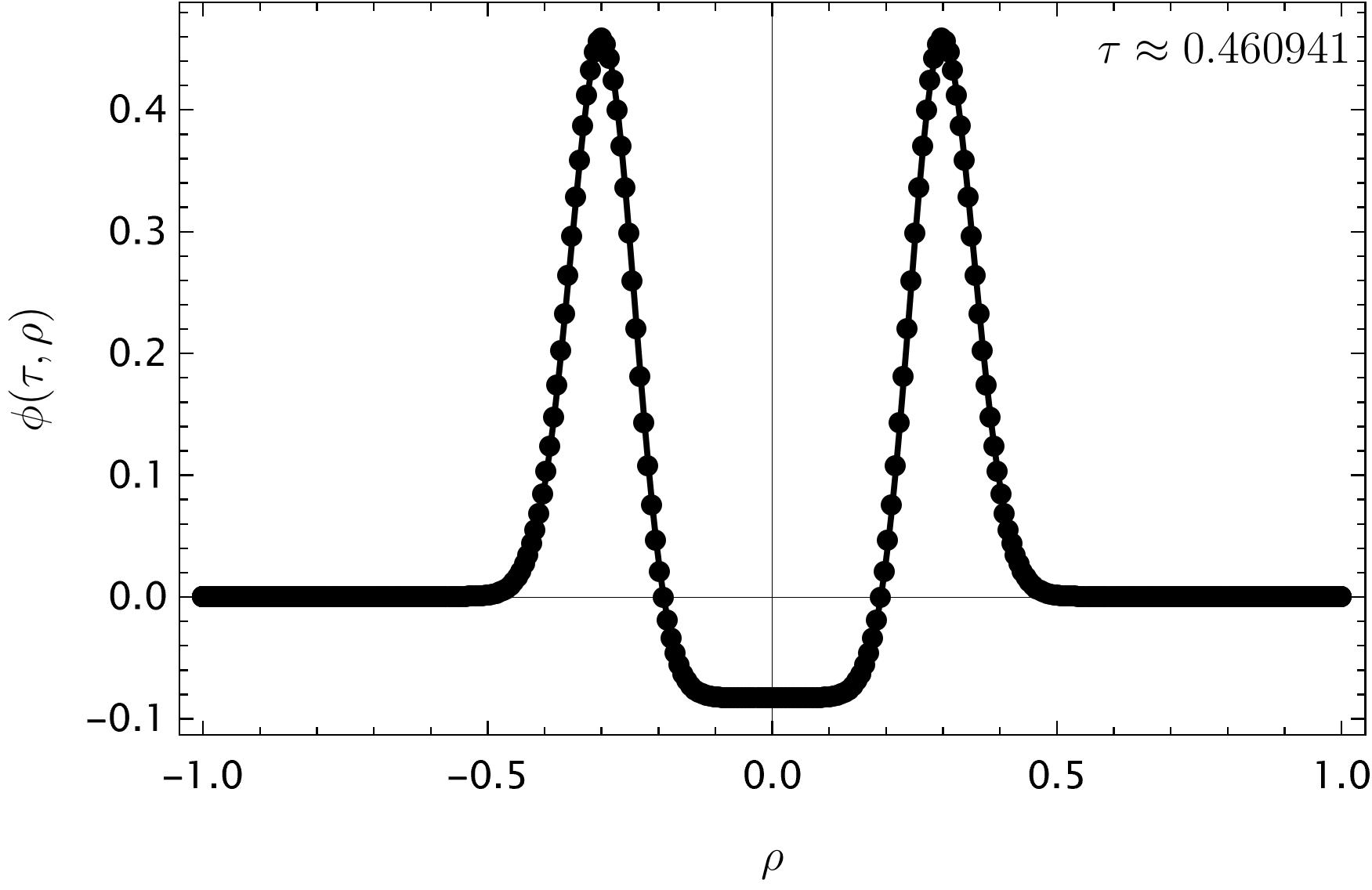}
    \quad
    \includegraphics[width=7cm]{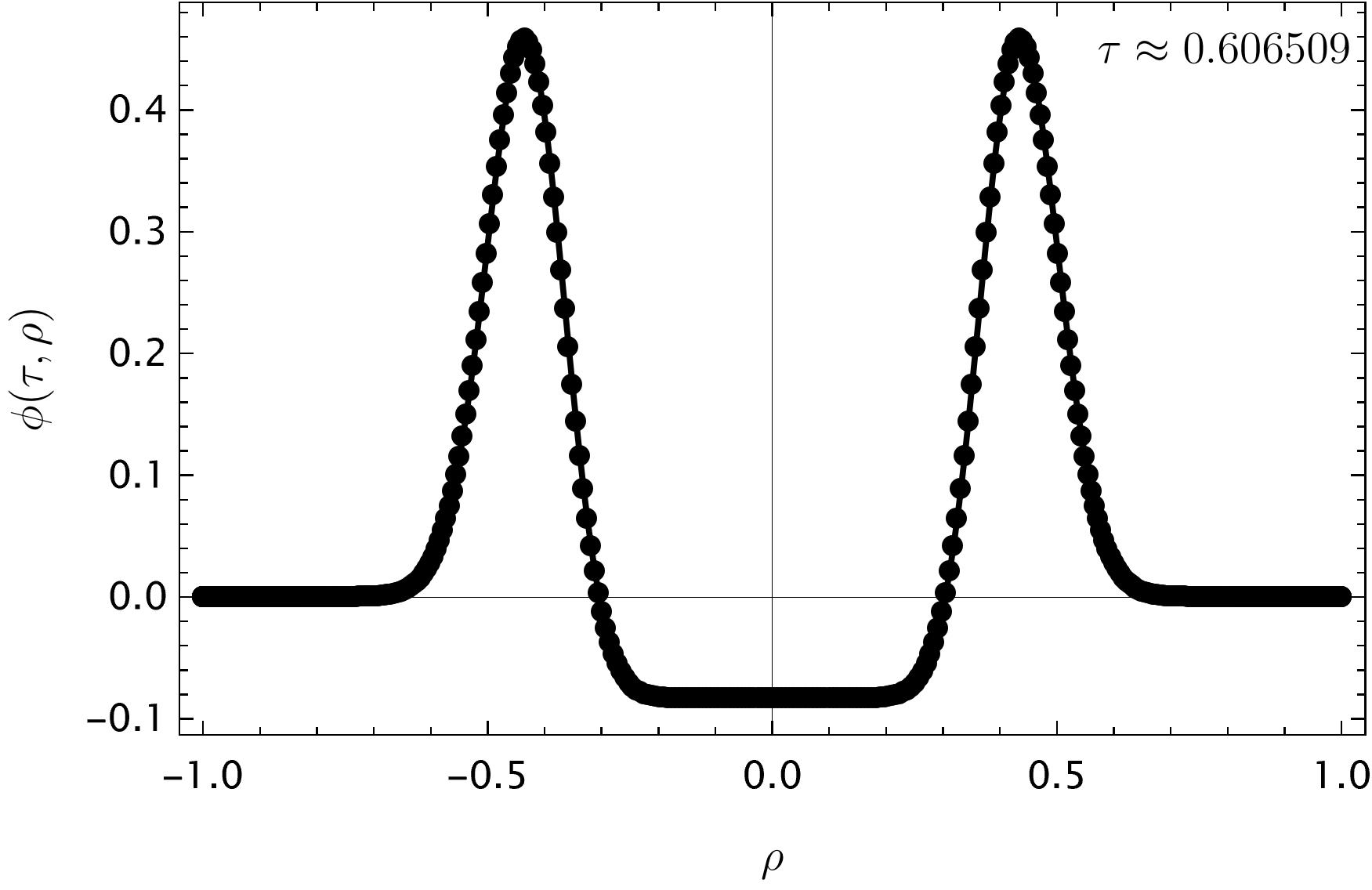}
    \quad
    \includegraphics[width=7cm]{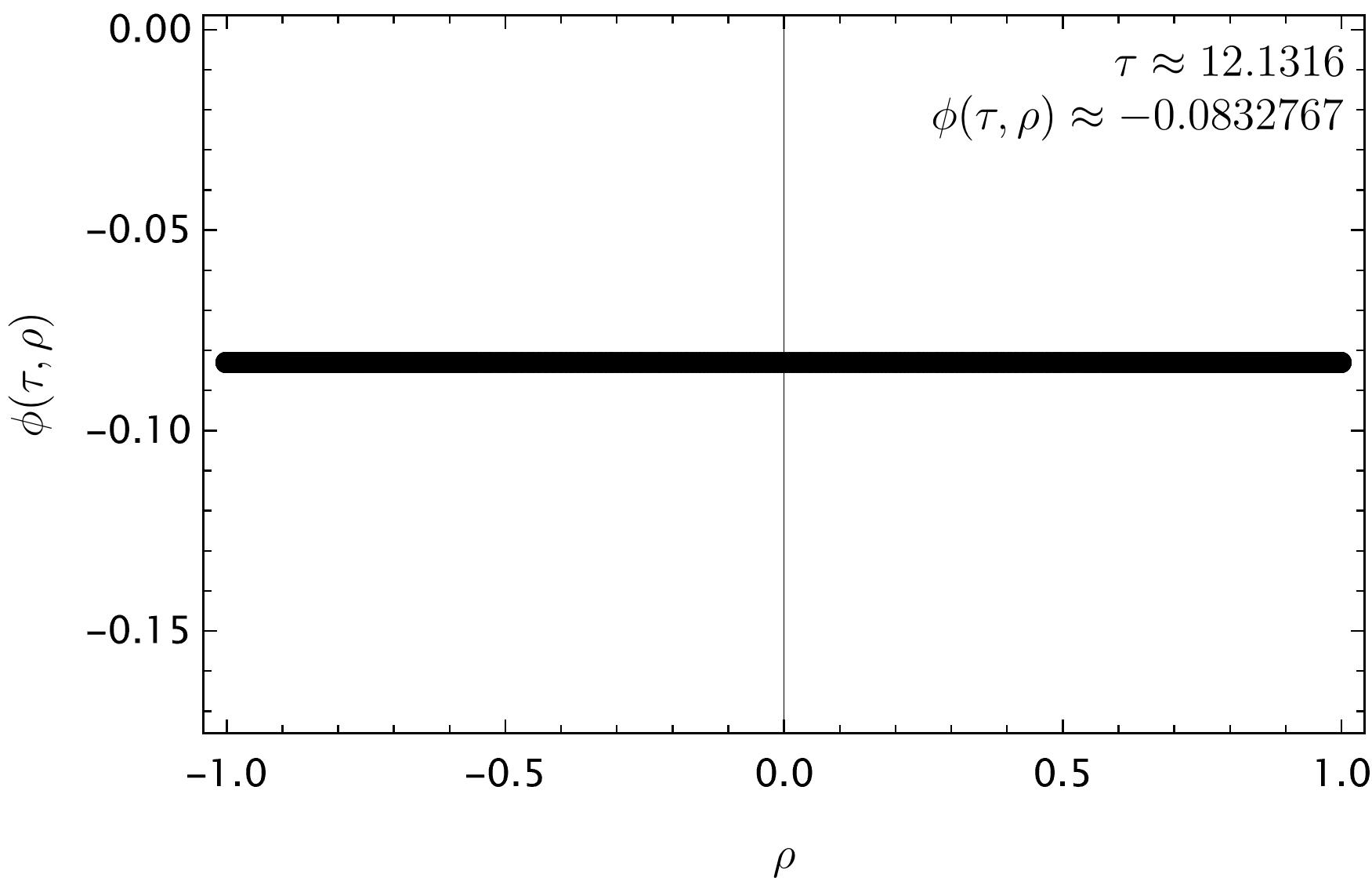}
    \caption{Numerical evolution of the \texttt{1+1D} wave equation defined by equation (\ref{HyperboloidalWaveEqn}) on a transformed hyperboloidal coordinate chart $t \rightarrow t(\tau,\rho), x \rightarrow x(\rho)$ numerically solved through the implementation of \texttt{IMTEX} Hermite time-integration scheme of order-4. As expected, outflow is observed, the initial pulse splits, travelling in opposite directions with equal field magnitudes until the solution leaves the numerical domain. However, unlike Figure \ref{fig:wave_tx_evolution}, we now observe a settling down to a negative constant, at around $\phi(\tau,\rho) \approx -0.0832767$, consistent with our theoretical predictions highlighted by equations \eqref{dAlembertsFormula}-\eqref{lim_tail} (last plot on the right).}
    \label{fig:wave_hyper_evolution}
\end{figure}

We check the accuracy of our numerical solution by computing the energy, which should be conserved while the pulse remains inside the numerical computational domain \cite{phdthesis-lidia, o2022conservativeX}. The energy is calculated as 
\begin{equation}
    E = \frac{1}{2}\int^{b}_{a}\bigg( \|\phi(t,x)\|^{2} + \|\mathcal{\psi}(t,x)\|^{2}  \bigg)\ dx, 
    \label{energy_def_cart}
\end{equation}
where here we use a Clenshaw-Curtis quadrature to integrate $x \in [a,b]$. The fractional energy error is calculated as 
\begin{equation}
    \frac{\delta E}{E_{0}} = \frac{E(t_{f}) - E_{t_{0}}}{E_{t_{0}}}, 
    \label{energy_fractional_error}
\end{equation}
and the numerical error as, 
\begin{equation}
    ||\Delta \phi(t,x)|| = \phi_{\mathrm{exact}}(t,x) - \phi_{\mathrm{numerical}}(t,x).  
    \label{numerical_error}
\end{equation}
\begin{figure}
    \centering
    \includegraphics[width=14cm]{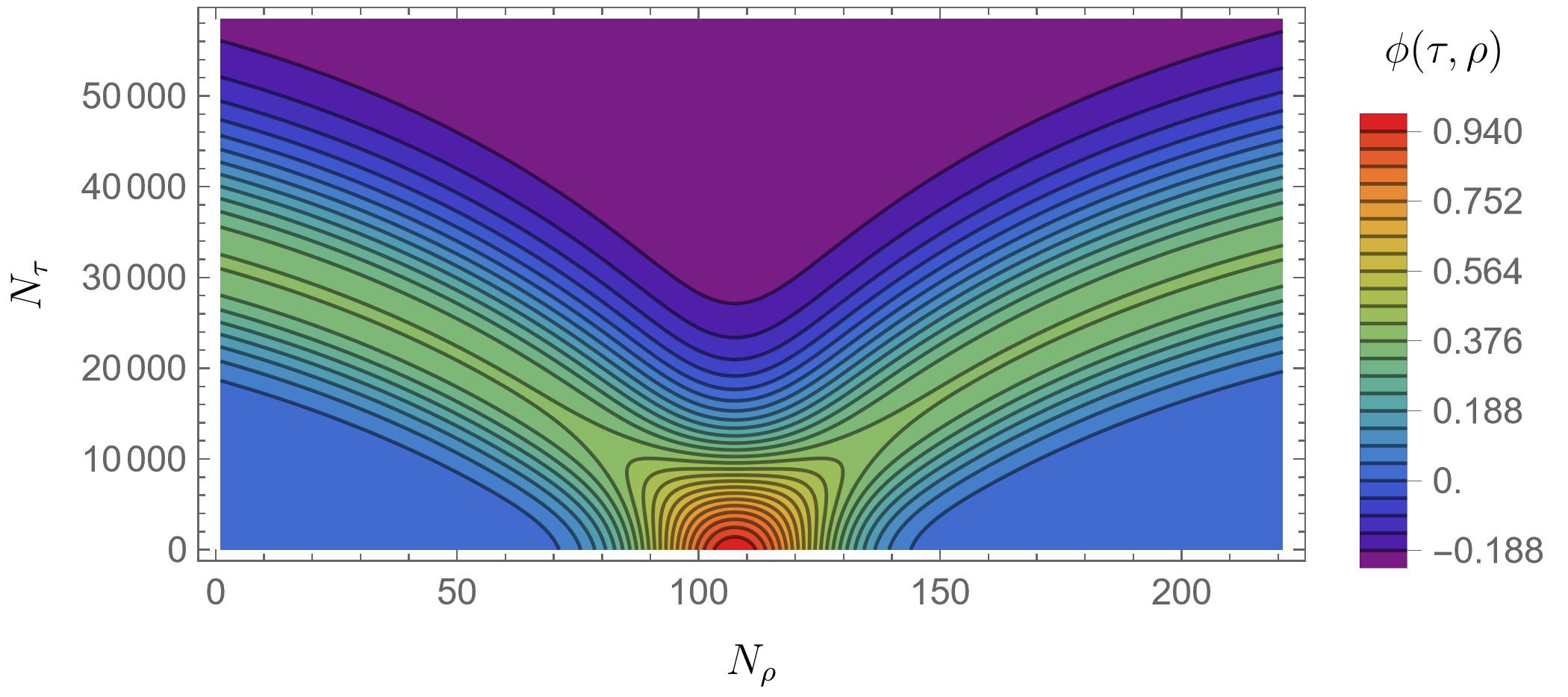}
    \caption{Contour plot further corroborating observations in Figure \ref{fig:wave_hyper_evolution}. Initially the gaussian pulse has maximum amplitude (highlighted in red) it then splits with equal amplitudes as the simulation runs its course. However, unlike in Figure \ref{fig:wave_tr_contour}, the simulation settles down to a negative constant (highlighted by the colour purple) at around $\phi(\tau,\rho) \approx -0.0832767$. }
    \label{fig:wave_hyper_contour_zen}
\end{figure}
Our numerical domain spans $x \in [-7,7]$ and we use \texttt{N=452} Chebyshev collocation nodes. Evolving our numerical solution we observe in Figure \ref{fig:wave_tx_evolution} outflow, as expected, for the entirety of the numerical simulation running time $t \approx [0., 25.4742]$. These results are further substantiated by the contour plot of Figure (\ref{fig:wave_tr_contour}). By the end of the simulation i.e at around $t=25.4742$ all that is left, is noise from the numerical framework used. To gauge the numerical error associated with the implementation of the numerical scheme, we calculate the L-infinity norm, $l_{\infty}$ of difference between the exact and numerical solutions as defined in eq.(\ref{numerical_error}). As given by the plot on the right of Figure \ref{fig:wave_tr_con_energy_linf} the solution remains accurate throughout the entire simulation. The accuracy of our simulations is further verified by computing the fractional error of the energy as defined in eq.(\ref{energy_fractional_error}). As we use \texttt{IMTEX} Hermite integration schemes we are able to conserve energy with near machine precision as showed in the left plot of Figure (\ref{fig:wave_tr_con_energy_linf}).

\subsection{Numerical solution to the wave equation in the ($t \rightarrow t(\tau,\rho), x \rightarrow x(\rho)$) hyperboloidal coordinate chart }

Our motivation is though to understand how the conversion of the Cartesian chart $(t,x)$ to the hyperboloidal chart described by equation (\ref{HyperboloidalWaveEqn}) i.e ($t \rightarrow t(\tau,\rho), x \rightarrow x(\rho)$) affects our numerical results. To that effect we implement the same numerical recipe as described above, however, now there will be no need to implement boundary conditions, as equation (\ref{HyperboloidalWaveEqn}) automatically enforces outflow. For initial data we make the following choice

\begin{subequations}
\begin{align}
    \label{conformal_wave_Psi_id}
    &\phi(0,\rho) = \exp\bigg( \frac{-\rho^{2}}{2 \ \Xi^{2}} \bigg) ,  \\
    &\partial_\tau\phi(0,\rho) = 0 ,  
    \label{conformal_wave_Pi_id}
\end{align}
\end{subequations}
where here $\Xi = 1/30$. For the sake of consistency, in the computation of the numerical solution to $\phi$ in the hyperboloidal chart $(\tau,\rho)$, we choose $N=452$ Chebyshev collocation nodes on a compact spatial domain spanning $\rho \in [-S,S] $ where $S=1$. Initially our simulation runs from $\tau \in [0,12.1306]$ with an optimal time-step of $\Delta \tau \approx 0.0000242613$. \\

\begin{table}
\centering
\begin{tabular}{||c c c c||} 
 \hline
    S & $\phi(\tau,\rho)|_{\tau = 20,000}$ & $l_{\infty}$ & $\Xi$  \\ [0.5ex] 
 \hline\hline
 1 & -0.0832767\textcolor{lightgrey}{3186260853} & $5.7 \times 10^{-7}$ & $\frac{1}{30}$ \\ 
 \hline
 4 & -0.0625484\textcolor{lightgrey}{6826822433} & $1.1 \times 10^{-7}$ & $\frac{1}{10}$\\
 \hline
 7 & -0.0357873\textcolor{lightgrey}{5053276409} & $2.8 \times 10^{-7}$ & $\frac{1}{10}$ \\
 \hline
 10 & -0.05007247\textcolor{lightgrey}{774450991} & $6.1 \times 10^{-8}$ & $\frac{1}{5}$ \\[1ex] 
 \hline
\end{tabular} 
\caption{Numerical solution $\phi(\tau,\rho)$ evaluated at relatively late times at around $\tau = 20, 000$ verifying the negative permanent displacement predicted by eqs.(\ref{DAlembertHyperboloidal}-\ref{lim_tail}) for different $S$ values. The numerical error given by the $l_{\infty}$ is calculated as per eq.(\ref{numerical_error}) where the exact solution is given by eqs.(\ref{DAlembertHyperboloidal}-\ref{lim_tail}).}
\label{tab_AllSsssssssssssss}
\end{table}

\begin{figure}
    \centering
    \includegraphics[width=6.8cm]{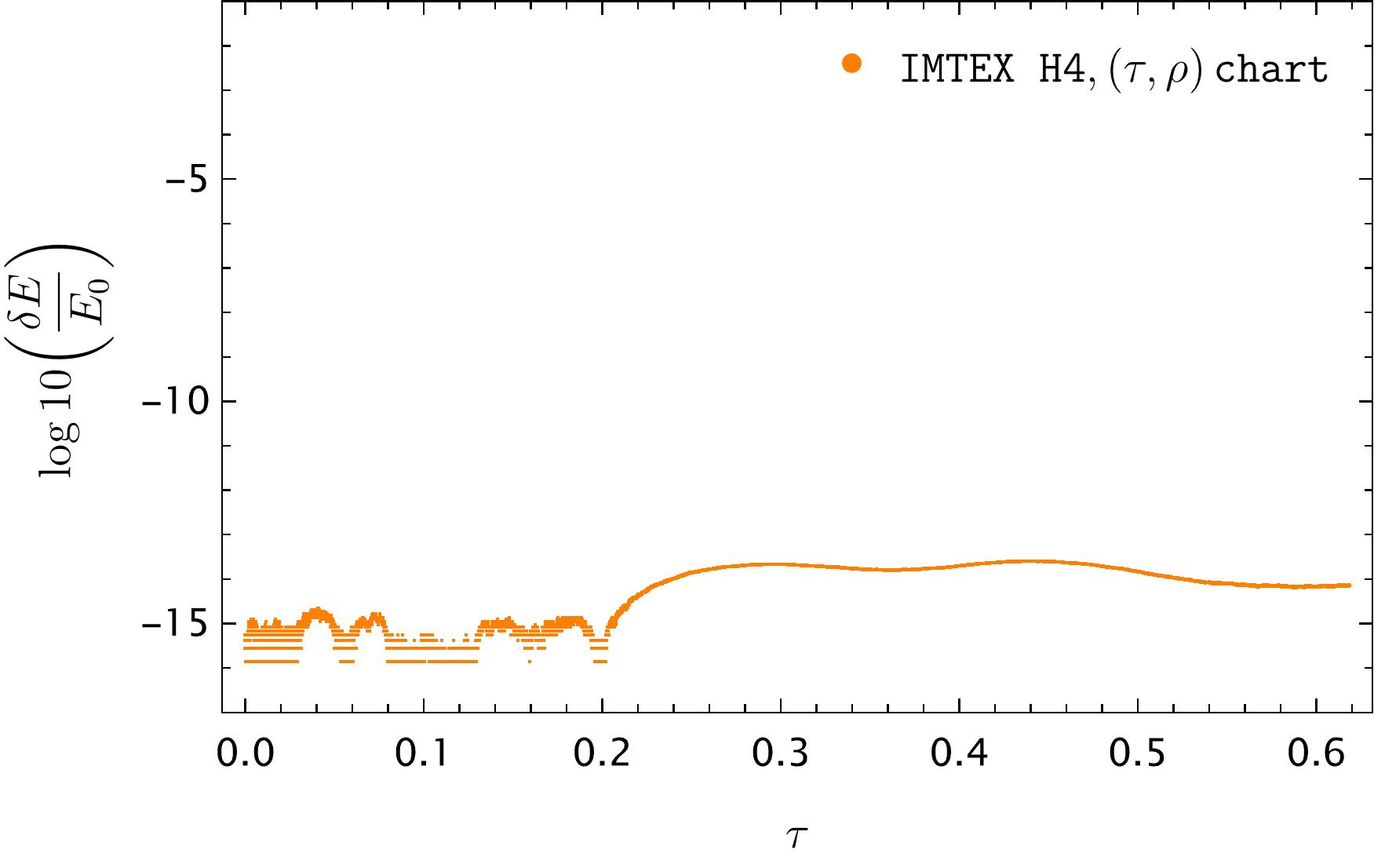}
    \quad 
    \includegraphics[width=8.3cm]{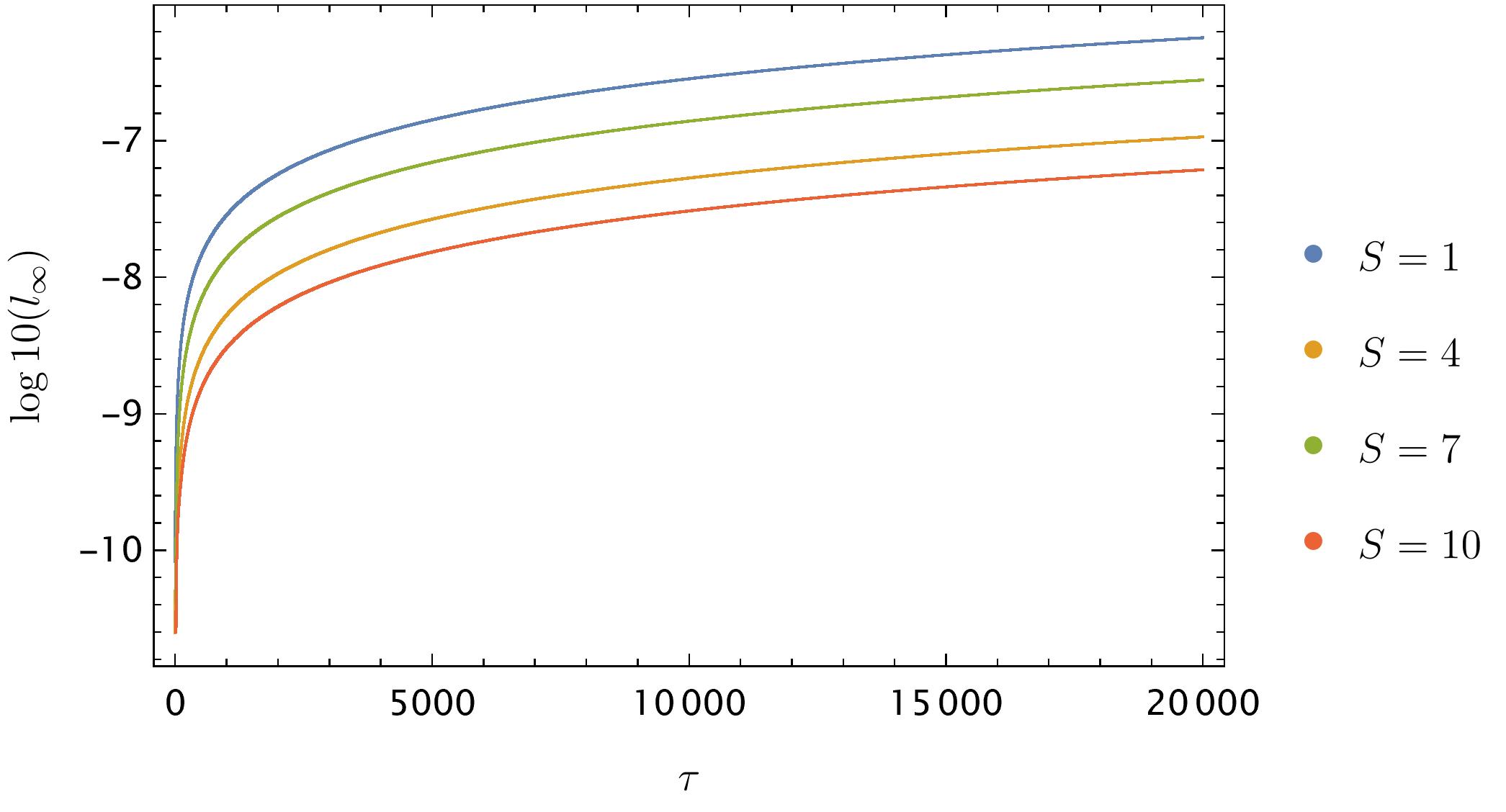}
    \caption{Numerical solution to the wave equation in the $(\tau,\rho)$ coordinate chart described by equation (\ref{HyperboloidalWaveEqn}). \textbf{Left}: fractional energy error of the numerical solution up to when the pulse is inside the computational domain. High accuracy is observed and energy is conserved both exactly and numerically \cite{phdthesis-lidia,da2024discotex, o2022conservativeX}. \textbf{Right}: numerical error between the numerical and exact solution predicted by eqs.(\ref{DAlembertHyperboloidal})-(\ref{lim_tail}) for different values of $S= {1,4,7,10}$ at very late times $\tau = [0, 20,000]$. As a compromise with simulation running times and unlike in Figure \ref{fig:wave_tr_con_energy_linf}, or the left plot here depicted, we significantly increased our time-step size to $\Delta \tau =0.02$. Our results are consistent with Figure \ref{fig:wave_tr_con_energy_linf}. }
    \label{fig:wave_hyper_con_energy_alsssss}
\end{figure}

As in the left plot of Figure \ref{fig:wave_tr_con_energy_linf}, here, in the left plot of Figure \ref{fig:wave_hyper_con_energy_alsssss}, we too observe conservation of energy to a fractional error near machine precision. However, as demonstrated in Figure \ref{fig:wave_hyper_evolution}, even though the initial behaviour is the same as that observed in the first two plots (from left to right) of Figure \ref{fig:wave_tx_evolution}, we now observe, Figure \ref{fig:wave_hyper_evolution}, a settling down to a negative constant, at around $\phi(\tau,\rho) \approx -0.0832767$, reflecting the negative permanent displacement predicted by our theoretical results highlighted in equations \eqref{DAlembertHyperboloidal}-\eqref{lim_tail}. This is further made clear by the contour plot given in Figure \ref{fig:wave_hyper_contour_zen} describing the numerical evolution of $\phi(\tau,\rho)$. To further assess the accuracy of our results we run our simulation for several values of $S = {1,4,7,10}$. Our interest is to verify that the value of the negative permanent displacement matches the theoretical prediction in eq.(\ref{lim_tail}), and hence it is necessary to significantly increase the simulation running time to $\tau = [0, 20,000]$. As a compromise with speed, we also significantly inflate the time-step size of our simulation to $\Delta \tau = 0.02 $. As it is observed in Table \ref{tab_AllSsssssssssssss}, the simulation remained accurate for all values investigated provided one ensured the initial gaussian pulse remained sufficiently compact, which can be easily achieved by calibrating the constant value $\Xi$. Both our theoretical and numerical results, as given by the plot on the right of Figure \ref{fig:wave_hyper_con_energy_alsssss}, are consistent with the accuracy of results obtained when solving the wave equation in the $(t,x)$. \\

It is important to note, that this behaviour is not exclusive to the particular chart given in \cite{Zen11a} but to any hyperboloidal chart. More generally, one would expect an analoguous effect in any coordinate transformation that modifies the time coordinate. For completion, we demonstrate this in \ref{Appendix_A} by implementing our numerical recipe to two different hyperboloidal slices.\\

Finally, the right plot in Figure \ref{fig:wave_hyper_con_energy_alsssss} shows the same fractional energy error as that obtained for the numerical solution in the $(t,x)$ chart, showing the merit of an hyperboloidal slice implementation. Not only does it further simplify our numerical implementation, it ensures numerical simulations remain accurate while automatically enforcing outflow behaviour. This advantage is of particular use when applied to problems such as those arising when modelling Extreme-Mass-Ratio-Inspirals (EMRIs) where accurate boundary conditions choices are less obvious and usually compromise the accuracy of the numerical solutions as discussed in \cite{da2024discotex, da2023hyperboloidal}. In Table 8 of \cite{da2024discotex} we give a brief outline of previous attempts within the community and it is noteworthy to highlight the merit of hyperboloidal approaches over traditional boundary conditions implementations on standard coordinate charts.

\section{Conclusions}
In this brief note we have obtained the hyperboloidal analogue of d'Alembert's solution to the wave equation in 1+1 dimensions. Our analysis has concentrated on a particular type of hyperboloidal foliation given in \cite{Zen11a}. The analogue expressions can be easily obtained for other families of hyperboloids such as those found in \cite{jaramillo2021pseudospectrum}. Moreover, the results can also be adapted to incorporate the spherically 3+1 dimensional wave equation ---for this one can make use of the method of spherical means to express the solution of the 3+1 equation in terms of a solution to the 1+1 problems. A detailed discussion of this solution goes beyond the illustrative purposes of this article. 

Our result brings to the fore the key role of explicit exact solutions in developing intuition into the nature of solutions to partial differential equations in non-standard physical and geometric settings.

\section*{Acknowledgements}
We would like to thank Rodrigo Panosso Macedo and Nelson Eir\'o for independently checking our numerical results in 2020-2021. We would also like to thank Anil Zenginoglu, Peter Diener, Alex Va\~{n}o-Vi\~{n}uales, Enno Harms and Sebastiano Bernuzzi for discussions and previous work. JAVK acknowledges the financial support of EPSRC grant \enquote{Geometric scattering methods for the conformal Einstein field equations}, EP/X012417/1.

\appendix

\section{Negative permanent displacement detection in other hyperboloidal coordinate charts numerical solutions to the wave equation} \label{Appendix_A}

For completion we now implement two different hyperboloidal slices as studied by \cite{jaramillo2021pseudospectrum} -- see also \cite{AnsMac14,AnsMac16,MacJarAns18,She22}, defined by the coordinate chart transformation, $\bar{t} \rightarrow (\bar{t}(\tau,\sigma), \bar{x} \rightarrow \bar{x}(\sigma))$. Following the notation of \cite{jaramillo2021pseudospectrum} we define, 
\begin{align}
    &\bar{t} = \tau - h(\sigma), \\
    &\bar{x} = g(\sigma), 
    \label{app_alt_slices}
\end{align}
where generically as shown in \cite{jaramillo2021pseudospectrum}, 
\begin{align}
    &\bar{t} =\frac{t}{\lambda}, \hspace{0.5cm} \bar{x} =\frac{r_{*}}{\lambda}, 
    \label{app_alt_slices_physical_qs}
\end{align}
with $\lambda$ to be specified further for each hyperboloidal slice we will study. $r_{*}$ is the tortoise coordinate given as $dr/dr_{\star} = f(r)$ where $f(r)=(1-2M/r)$, in an extended domain of $r_{\star} = [\infty,\infty]$. Furthermore, $h(\sigma)$ is the height function implementing the hyperboloidal slicing and $g(\sigma)$ is the spatial compactification bringing $\bar{x} \in [-\infty, \infty]$ down to a compact domain of $\sigma = [a,b]$. 

The first slice we will study, which we will here call Slice A uses compactified hyperboloids given by Biz\'on-Mach coordinates and maps $\mathbb{R}$ to $]-1,1[$ via, 
\begin{align}
    &\bar{t} = \tau - h(\sigma) = \tau - \frac{1}{2}\ln{(1-\sigma^{2})}, \\
    &\bar{x} = \rm{arctanh}(\sigma),
    \label{sliceA}
\end{align}
with $\lambda = 1$ and $\bar{x} \in [-\infty,\infty]$ being compactified to the domain of $[a,b]$ where $a=-1, b=1$. The evolution operators $\textbf{L}_{1}, \textbf{L}_{2}$ as defined in equations (\ref{Genric_matrix_op}-\ref{phi_psi_ODEs}) now transform to, 
\begin{subequations}
\begin{align}
    \label{ch3_sb2_diffOperators_varepsilon_rpmf}
    & \tilde{\varepsilon}(\sigma) = - 2 \sigma , \\
    \label{ch3_sb2_diffOperators_varrho_rpmf}
    &\tilde{\varrho}(\sigma) =  -1 , \\
    \label{ch3_sb2_diffOperators_chi_rpmf}
    &\tilde{\chi}(\sigma) = (1 - \sigma^{2}),\\
    \label{ch3_sb2_diffOperators_potential_rpmf} 
    &\tilde{\mathcal{V}}(\sigma) = 0, \\
    &\tilde{\iota}(\sigma) = - 2 \sigma. 
    \label{ch3_sb2_diffOperators_iota_rodrigo_flat}
\end{align}
\end{subequations}
For initial data, we inject the following gaussian pulse
\begin{align}
    \label{conformal_wave_Psi_id_rpm}
    &\phi(0,\sigma) = \exp\bigg({-a\bigg( - \frac{1}{2} \log{(1-\sigma^{2})} - \arctan{(\sigma)} \bigg)^{2}\bigg)}, \\
     &\partial_{\tau}\phi(0,\sigma) = 0 ,
    \label{conformal_wave_Pi_id_rpm}
\end{align}
where $a=350$. The second slice, hereafter slice B, commonly referred to as \textit{minimal gauge} is attained via the transformations, 
\begin{align}
    &\bar{t} = \tau - h(\sigma) = \tau - \frac{1}{2}\bigg(\ln{\sigma} + \ln{(1-\sigma)} - \frac{1}{\sigma} \bigg),\\
    &\bar{x} =  \frac{1}{2}\bigg(\ln{(1-\sigma)} + \frac{1}{\sigma} - \ln{\sigma}\bigg),
    \label{sliceB}
\end{align}
where here $\lambda = 4M$ and $\bar{x} = r_{*}/\lambda $, and thus $\bar{x} \in [-\infty,\infty]$ is mapped to the compactified domain of $[a,b]$ where $a=0, b=1$. 

\begin{figure}
    \centering
    \includegraphics[width=12cm]{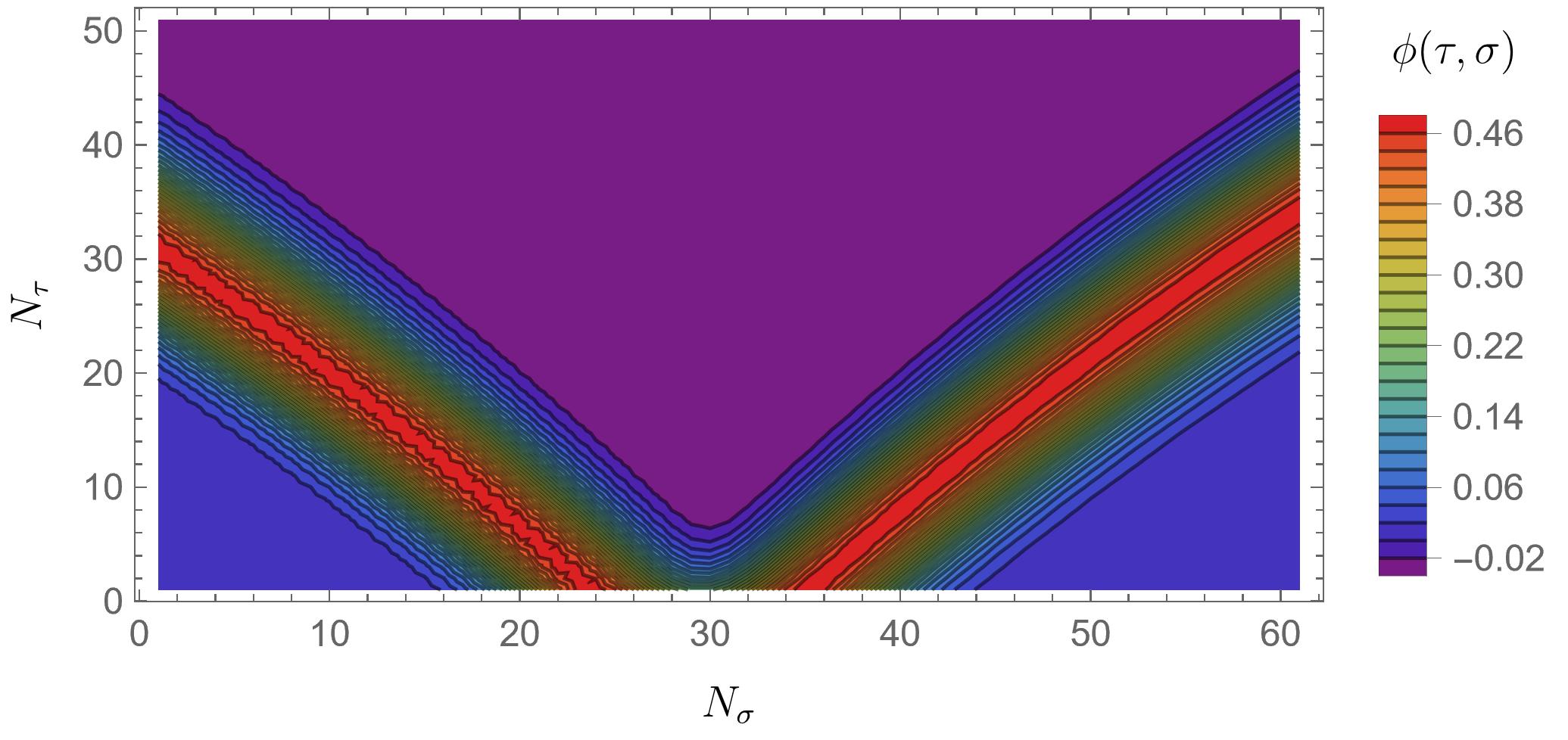}
    \quad 
    \includegraphics[width=12cm]{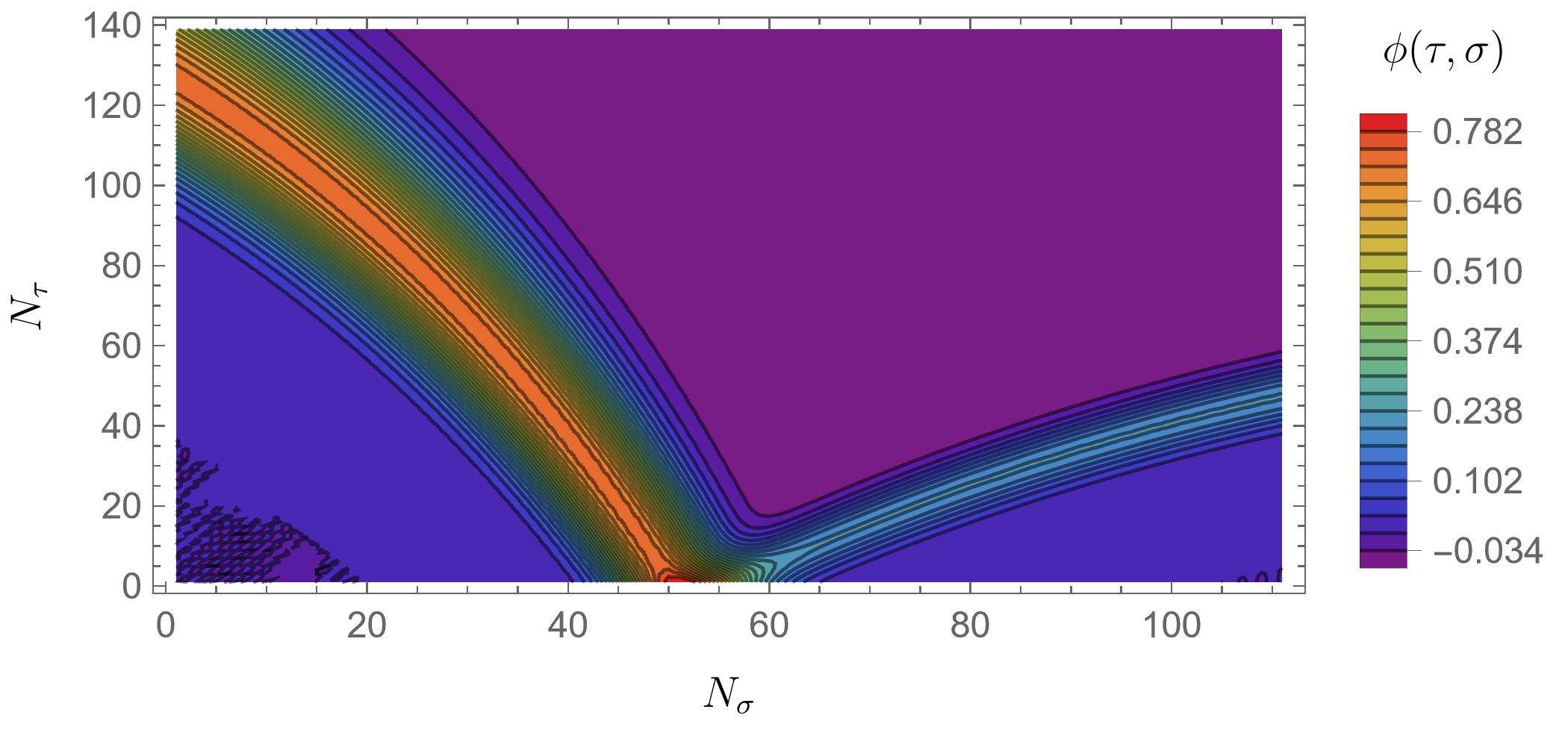}
    \quad 
    \includegraphics[width=12cm]{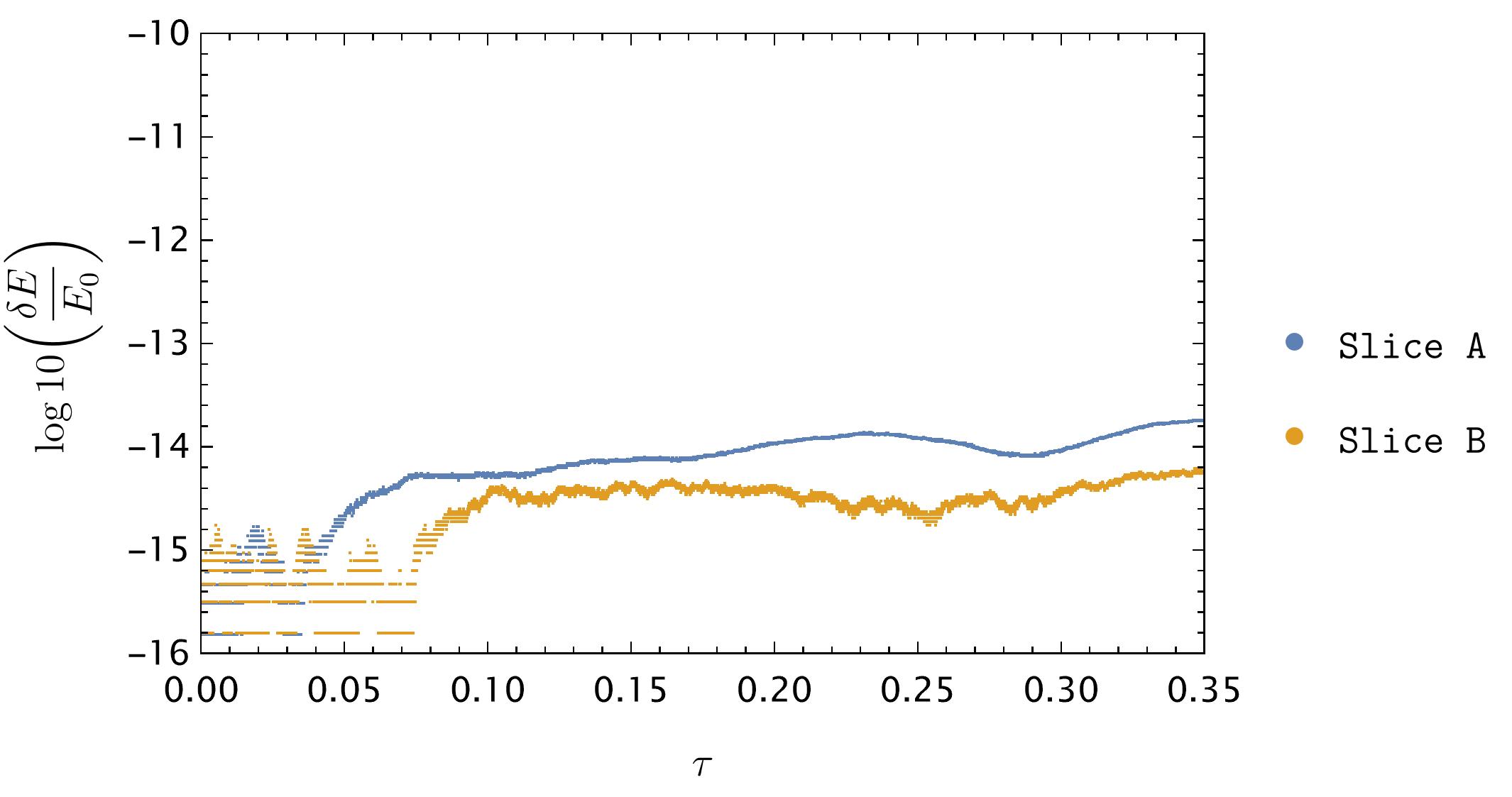}
    \caption{Numerical solution to the wave equation in two hyperboloidal  $(\tau,\sigma)$ coordinate charts. \textbf{Top, Middle}: contour plots showing the negative permanent displacement of two different slices, slice A and slice B of a value of $\phi(\tau,\sigma) \approx  -0.0474046$  and $\phi(\tau,\sigma) \approx -0.0672598$, respectively. \textbf{Bottom}: fractional energy error of the numerical solutions of Slice A and Slice B up to when the pulse is inside the computational domain. High accuracy is observed and energy is conserved both exactly and numerically \cite{ phdthesis-lidia, da2024discotex, o2022conservativeX} and is consistent with the result observed in Figures \ref{fig:wave_tr_con_energy_linf},\ref{fig:wave_hyper_con_energy_alsssss}. }
    \label{fig:wave_hyper_con_energy_rpm_slices}
\end{figure}

For the second slice, slice B, commonly referred to as the \textit{minimal gauge}, the evolution operators are exactly as defined by equations (27)-(31) of \cite{da2023hyperboloidal} with the difference that in this work $\tilde{\mathcal{V}}(\sigma)=0$. For initial data for slice B, we pick, 
\begin{subequations}
\begin{align}
    \label{conformal_wave_Psi_id}
    &\phi(0,\sigma) = \exp\bigg( \frac{-(\sigma - 0.5)^{2}}{2 \ \Xi} \bigg) ,  \\
    &\partial_\tau\phi(0,\sigma) = 0 ,  
    \label{conformal_wave_Pi_id_rpm_minimal}
\end{align}
\end{subequations}
where $\Xi = 1/3000$. For the numerical simulation of slice A, we use $N=256$ Chebyshev collocation nodes on spatial domain spanning over $\sigma \in [-1,1]$, a time-step size of $\Delta \tau = 0.0000379449 $ over a time interval of $\tau = [0, 18.974]$. In the numerical simulation of slice B, we use $N=256$ Chebyshev collocation nodes on spatial domain spanning over $\sigma \in [0,1]$, a time-step size of $\Delta \tau \approx 0.00001897247 $ over a time interval of $\tau = [0, 18.974]$. As we can see from the first two plots, Top and Middle of Figure \ref{fig:wave_hyper_con_energy_rpm_slices} a negative permanent displacement is observed at late times of $\phi(\tau,\sigma) \approx  -0.0474046$  and $\phi(\tau,\sigma) \approx -0.0672598$ for slice A and slice B, respectively. In the latter, and unlike in the other two slices investigated in this work, the splitting of the gaussian pulse is not symmetric, this is due the fact the slicing was derived for curved spacetime and designed to improve the regularity of Schwarzschild (-like) cases \cite{jaramillo2021pseudospectrum}. As with Figures \ref{fig:wave_tr_con_energy_linf},\ref{fig:wave_hyper_con_energy_alsssss} energy conservation is observed. \footnote{We thank Rodrigo Panosso Macedo for having suggested and numerically independently verified our results with these slices \cite{rodrigo_pc}.} Checking the numerical error would require the derivation of an analogous hyperboloidal d'Alembert equation as showed in eqs.(\ref{DAlembertHyperboloidal}-\ref{lim_tail}) and is out of scope of this work. 

\clearpage
\section*{References}
\bibliographystyle{unsrt}

\begin{thebibliography}{10}

\bibitem{CFEBook}
J.~A. {Valiente Kroon}.
\newblock {\em Conformal Methods in General Relativity}.
\newblock Cambridge University Press, 2016.

\bibitem{Zen24}
A.~Zenginoglu.
\newblock Hyperbolic times in minkowski space.
\newblock In {\tt arXiv:2404.01528}, 2024.

\bibitem{Fri83}
H.~Friedrich.
\newblock Cauchy problems for the conformal vacuum field equations in general relativity.
\newblock {\em Comm. Math. Phys.}, 91:445, 1983.

\bibitem{Fri86b}
H.~Friedrich.
\newblock On the existence of n-geodesically complete or future complete solutions of {E}instein's field equations with smooth asymptotic structure.
\newblock {\em Comm. Math. Phys.}, 107:587, 1986.

\bibitem{Zen11b}
A.~Zenginoglu.
\newblock A geometric framework for black hole perturbations.
\newblock {\em Phys. Rev. D}, 83:127502, 2011.

\bibitem{ZenKid10}
A.~Zenginoglu and L.~E. Kidder.
\newblock Hyperboloidal evolution of test fields in three spatial dimensions.
\newblock {\em Phys. Rev. D}, 81:124010, 2010.

\bibitem{GauVanHilBos21}
S.~Gautam, A.~Va\ {n}o Vi\~{n}uales, D.~Hilditch, and S.~Bose.
\newblock Summation by parts and truncation error matching on hyperboloidal slices.
\newblock {\em Phys. Rev. D}, 103:084045, 2021.

\bibitem{Zen11a}
A.~Zenginoglu.
\newblock Hyperboloidal layers for hyperbolic equations on unbounded domains.
\newblock {\em J. Comput. Phys.}, 230:2286, 2011.

\bibitem{HagLau07}
T.~Hagstrom and S.~Lau.
\newblock Radiation boundary conditions for maxwell's equations" a review of accurate time-domain formulations.
\newblock {\em J. Comp. Math.}, 25:305, 2007.

\bibitem{Lau04}
S.~Lau.
\newblock Radiation boundary kernels for time-domain wave propagation on black holes: thoery and numerical mathods.
\newblock {\em J. Comp. Phys.}, 199:376, 2004.

\bibitem{Lau05}
S.~Lau.
\newblock Analytic structure of radiation boundary kernels for blackhole perturbations.
\newblock {\em J. Math. Phys.}, 46:102503, 2005.

\bibitem{FieLau15}
S.~Lau.
\newblock Fast evaluation of far-field signals for time-domain wave propagation.
\newblock {\em J. Sci. Comp.}, 64:647, 2015.

\bibitem{GroKel95}
M.~J. Grote and J.~B. Keller.
\newblock On nonreflecting boundary conditions.
\newblock {\em J. Comp. Phys.}, 122:231, 1995.

\bibitem{Eva98}
L.~C. Evans.
\newblock {\em Partial Differential Equations}.
\newblock American Mathematical Society, 1998.

\bibitem{Joh91}
F.~John.
\newblock {\em Partial differential equations}.
\newblock Springer, 1991.

\bibitem{eharmsmasters2012}
Enno Harms.
\newblock Numerical solution of the 2+1 teukolsky equation on a hyperboloidal foliation of the kerr spacetime.
\newblock Master's thesis, Friedrich-Schiller-Universit\"at Jena, Physikalisch-Astronomische Fakult\"at, Deutschland, 2012.

\bibitem{rodrigo_pc}
Rodrigo~Panosso Macedo.
\newblock Fully-spectral minimal gauge time-domain code.
\newblock {\em Private Communication}, 2021.

\bibitem{phdthesis-lidia}
L.~J. {Gomes Da Silva}.
\newblock {\em Numerical Algorithms for the modelling of {EMRI}s in the time domain}.
\newblock {PhD} dissertation, QMUL University, School of Mathematical Sciences, 2023.

\bibitem{da2024discotex}
Lidia J~Gomes Da~Silva.
\newblock Discotex: Discontinuous collocation and implicit-turned-explicit (imtex) integration symplectic, symmetric numerical algorithms with higher order jumps for differential equations with numerical black hole perturbation theory applications.
\newblock {\em arXiv preprint arXiv:2401.08758}, 2024.

\bibitem{o2022conservativeX}
Michael~F O'Boyle, Charalampos Markakis, Lidia J~Gomes Da~Silva, Nelson Eir\'o, Rodrigo~Panosso Macedo, and Juan A~Valiente Kroon.
\newblock Conservative evolution of black hole perturbations with time-symmetric numerical methods.
\newblock {\em arXiv preprint arXiv:2210.02550}, 2022.

\bibitem{da2023hyperboloidal}
Lidia~J Gomes Da~Silva, Rodrigo~Panosso Macedo, Jonathan~E Thompson, Juan A~Valiente Kroon, Leanne Durkan, and Oliver Long.
\newblock Hyperboloidal discontinuous time-symmetric numerical algorithm with higher order jumps for gravitational self-force computations in the time domain.
\newblock {\em arXiv preprint arXiv:2306.13153}, 2023.

\bibitem{jaramillo2021pseudospectrum}
Jos{\'e}~Luis Jaramillo, Rodrigo~Panosso Macedo, and Lamis Al~Sheikh.
\newblock Pseudospectrum and black hole quasinormal mode instability.
\newblock {\em Physical Review X}, 11(3):031003, 2021.

\bibitem{AnsMac14}
M.~Ansorg and R.~P. Macedo.
\newblock Axisymmetric fully spectral code for hyperbolic equations.
\newblock {\em J. Comp. Phys.}, 276:357, 2014.

\bibitem{AnsMac16}
M.~Ansorg and R.~P. Macedo.
\newblock Spectral decomposition of black-hole perturbations on hyperboloidal slices.
\newblock {\em Phys. Rev. D}, 93:124016, 2016.

\bibitem{MacJarAns18}
R.~P. Macedo, J.~L. Jaramillo, and M.~Ansorg.
\newblock Axisymmetric fully spectral code for hyperbolic equations.
\newblock {\em Phys. Rev. D}, 98:124005, 2018.

\bibitem{She22}
L.~Al Sheik.
\newblock {\em Scattering resonances and Pseudospectrum : stability and completeness aspects in optical and gravitational systems}.
\newblock PhD thesis, Institut de Math\'ematiques de Bourgogne, Dijon, France, 2022.

\end{thebibliography}

\end{document}